\documentclass[fleqn,usenatbib]{mnras}

\usepackage{newtxtext,newtxmath,epsfig} 

\usepackage[T1]{fontenc}
\usepackage{ae,aecompl}

\DeclareRobustCommand{\VAN}[3]{#2}
\let\VANthebibliography\thebibliography
\def\thebibliography{\DeclareRobustCommand{\VAN}[3]{##3}\VANthebibliography}

\usepackage{graphicx}	
\usepackage{amsmath}	
\usepackage[dvipsnames,svgnames,x11names]{xcolor}
\usepackage[multiple]{footmisc}
\usepackage{multirow}
\usepackage{tcolorbox}
\usepackage{arydshln}
\usepackage{wasysym}

\usepackage{ulem}   

\definecolor{lightcoral}{rgb}{0.94, 0.5, 0.5}

\definecolor{mpc}{rgb}{0.2, 0.2, 0.8}
\definecolor{GR}{rgb}{0.8, 0.2, 0.2}

\defcitealias{Torbaniuk:21}{Paper~I}

\title[SMBH growth in low-z galaxies]{Probing supermassive black hole growth and its dependence on stellar mass and star-formation rate in low-redshift galaxies}

\author[O. Torbaniuk et al.]{O. Torbaniuk$^{1,2}$\thanks{E-mail: olena.torbaniuk@gmail.com},
M. Paolillo$^{3,4,5}$,
R. D'Abrusco$^{6}$,
C. Vignali$^{1,2}$,
A. Georgakakis$^{7}$,
F. J. Carrera$^{8}$,
F. Civano$^{9}$
\\
$^{1}$Department of Physics and Astronomy `Augusto Righi', University of Bologna, via Piero Gobetti 93/2, I-40129 Bologna, Italy\\
$^{2}$INAF -- Osservatorio di Astrofisica e Scienza dello Spazio di Bologna, Via Gobetti 101, I-40129 Bologna, Italy\\
$^{3}$Department of Physics, University of Napoli Federico II, via Cinthia 9, 80126, Napoli, Italy\\
$^{4}$INAF -- Osservatorio Astronomico di Capodimonte, via Moiariello 16, 80131, Napoli, Italy\\
$^{5}$INFN -- Sezione di Napoli, via Cinthia 9, 80126, Napoli, Italy\\
$^{6}$Harvard-Smithsonian Center for Astrophysics, 60 Garden Street, Cambridge, MA, 02138, USA\\
$^{7}$Institute for Astronomy \& Astrophysics, National Observatory of Athens, V. Paulou \& I. Metaxa, Athens, 11532, Greece\\
$^{8}$Instituto de Física de Cantabria (CSIC-Universidad de Cantabria), Avenida de los Castros, 39005 Santander, Spain\\
$^{9}$Astrophysics Science Division, NASA Goddard Space Flight Center, Greenbelt, MD 20771, USA
}

\date{Accepted XXX. Received YYY; in original form ZZZ}

\pubyear{2023}

\begin{document}
\label{firstpage}
\pagerange{\pageref{firstpage}--\pageref{lastpage}}
\maketitle

\begin{abstract}

We present an improved study of the relation between supermassive black hole growth and their host galaxy properties in the local Universe ($z < 0.33$). To this end, we build an extensive sample combining spectroscopic measurements of star-formation rate (SFR) and stellar mass from Sloan Digital Sky Survey, with specific Black Hole accretion rate (sBHAR, $\lambda_{\mathrm{sBHAR}} \propto L_{\rm X}/\mathcal{M}_{\ast}$) derived from the \textit{XMM-Newton} Serendipitous Source Catalogue (3XMM-DR8) and the \textit{Chandra} Source Catalogue (CSC\,2.0).
We find that the sBHAR probability distribution for both star-forming and quiescent galaxies has a power-law shape peaking at $\log\lambda_{\mathrm{sBHAR}}\sim -3.5$ and declining toward lower sBHAR in all stellar mass ranges. This finding confirms the decrease of AGN activity in the local Universe compared to higher redshifts. We observe a significant correlation between $\log\,\lambda_{\rm sBHAR}$ and $\log\,{\rm SFR}$ in almost all stellar mass ranges, but the relation is shallower compared to higher redshifts, indicating a reduced availability of accreting material in the local Universe. At the same time, the BHAR-to-SFR ratio for star-forming galaxies strongly correlates with stellar mass, supporting the scenario where both AGN activity and stellar formation  primarily depend on the stellar mass via fuelling by a common gas reservoir. Conversely, this ratio remains constant for quiescent galaxies, possibly indicating the existence of the different physical mechanisms responsible for AGN fuelling or different accretion mode in quiescent galaxies. 
\end{abstract}



\section{Introduction}


The growth of galaxies (via stellar formation processes) and of the supermassive black hole (SMBH) at their centers (via mass-accretion potentially triggering an Active Galactic Nucleus or AGN) appear to proceed coherently over cosmic times. As was suggested by several studies, the global cosmological star-formation rate and AGN accretion rate show similar evolution with redshift, reaching a peak at redshift $z \sim 1-3$ and declining rapidly towards more recent cosmic times \citep{Delvecchio:14, Madau:14, Aird:15, Malefahlo:22, DSilva:23}. At the same time, the mass of central SMBHs seem to be tightly correlated with the properties of their host galaxies (e.g. stellar velocity dispersion, bulge mass, total stellar mass; see \citealt{Ferrarese:00, Gultekin:09, McConnel:13, Kormendy:13, Reines:15, Shankar:16, Gonzalez:22, LiJ:23, Poitevineau:23, Sahu:23}). However, the physical origin of such correlations is still poorly understood. 
 
Since the rate of star formation on galactic scales is directly related to the availability of cold gas (and its efficiency in forming stars, \citealt{Bigiel:08, Peng:14, Catinella:18}), it is reasonable to speculate that the nuclear activity can be fed by the same gas reservoir being accreted onto the central SMBH. This scenario also agrees with studies showing that moderate-to-high luminosity AGN predominately reside in galaxies with higher star-formation rates \citep{Merloni:10, Rosario:13, Heinis:16, Aird:18, Stemo:20, Torbaniuk:21}. Still, as AGN accretion operates typically on smaller spatial scales than star formation, the confirmation of the presence of the common gas reservoir for SMBH accretion and star formation requires a deeper understanding of the mechanism responsible for gas transportation from the galaxy outskirts all the way to their centers. Individual observations of the closest galaxies and hydrodynamical simulations suggest that such mechanism can be provided by large-scale gravitational torques formed by disk instabilities or by major mergers and minor interactions among galaxies \citep{Fathi:06, Hopkins:10, Fischer:15, Storchi-Bergmann:19, Quai:23}. In addition, the feedback provided by stellar evolution and AGN activity may be pivotal in controlling the amount and distribution of cold gas and consequently alter AGN growth. For instance, the stellar feedback can significantly affect nuclear accretion reducing the gas supply by star formation or, on the contrary, enhancing it through the turbulence injection produced by supernova explosions and/or strong wind from massive stars \citep{Schartmann:09, Hopkins:16, Byrne:23}. At the same time,  accretion onto the SMBH can generate energetic outputs in the form of electromagnetic radiation (i.e. radiative or quasar mode) or powerful jets (i.e. radio or jet mode) depending on the accretion efficiency ($>1$ and $\ll 1$\,per\,cent Eddington, respectively; \citealt{Fabian:12, Heckman:14}). Hence, interacting with the gas in the host galaxy through radiation pressure that produces a powerful wind (quasar mode) or an outflow of relativistic particles (jet mode) AGN can suppress the formation of stars by heating and/or blowing the cold gas away from the galaxy or, vice versa, trigger it by compressing dense clouds in the interstellar medium, \citealt{Schawinski:09, Ishibashi:12, Zubovas:13, Leslie:16, Combes:17, Fiore:17, Park:23, Ferrara:23}). In addition to being able to quench star formation in the host galaxy, AGN feedback can also significantly reduce its own accretion, thus resulting in self-regulation of the nuclear activity \citep{Fabian:12, Paspaliaris:23}. 

However, all mechanisms of gas transportation discussed above are not able to produce the continuous and regular gas flow required to feed the SMBH and therefore, they trigger AGN activity in a much more stochastic manner compared to the smooth and coherent process of star formation. Such stochasticity is usually observed as the variability of the AGN activity, which may lead to the change of AGN luminosity by orders of magnitude on intermediate-long timescales of $\sim10^{5-7}$\,yr (see \citealp{Mullaney:12a, Aird:13, King:15, Sartori:18}). Thus,  observations of any individual AGN are probing only a fraction of its activity cycle and therefore do not allow to  study directly the connection between the growth of the central SMBH and the overall galaxy properties; this agrees with recent observations swing that galaxies with similar stellar masses and SFR contain AGN with a very broad range of  accretion rates \citep{Bongiorno:12, Aird:12, Georgakakis:14, Aird:18, Torbaniuk:21}. 
Therefore, the proper investigation of the relation between galaxy properties and AGN activity requires the study of representative samples of the whole AGN population. The availability of large statistical samples of galaxies allows to probe the AGN activity over cosmological scales through the BHAR probability density function, i.e. the probability of a galaxy with certain properties (e.g. stellar mass, SFR, morphological type) to host a SMBH accreting at a certain rate. According to recent works, the BHAR probability function seems to have a power-law shape with an exponential cut-off at high BHAR and with flattening or even decreasing toward low accretion rates \citep{Aird:12, Bongiorno:12, Aird:18}.

Studies of such correlation between AGN activity and host galaxy properties require careful separation of the nuclear and stellar emission. The detection of X-ray radiation produced by the innermost regions of active nuclei is an efficient method for probing SMBH accretion over a wide range of redshifts. Moreover, it allows us to access even relatively low-luminosity AGN in the local Universe, where the identification in the optical and infrared bands is not trivial due to the contamination of the host galaxy \citep{Brandt:05, Alexander:12, Merloni:16}. However, due to the lack of deep, uniform, wide-areas X-ray surveys most studies on the co-evolution between galaxies and their central SMBH have focused on intermediate/high redshift from $z \sim 0.25$ up to $z \approx 4.0$ \citep{Chen:13, Rosario:13, Delvecchio:15, Rodighiero:15, Aird:18, Aird:19, Stemo:20, Spinoglio:22, Pouliasis:22} and therefore, the link between BH accretion rate and SFR (or stellar mass) have been investigated mainly for the moderate- and high-luminosity AGNs. Understanding the BHAR--SFR relation also in local galaxies, in addition to probing the bulk of the accreting BH population, i.e. low-to-moderate luminosity AGN, also provides information on the physics of low efficiency SMBH accretion, which is difficult to trace at higher redshifts. Moreover, the population of galaxies with fading star formation (i.e. quiescent galaxies) in the local Universe is a crucial laboratory to evaluate the role of AGN activity in star-formation suppression and to explore the alternative mechanisms of AGN fuelling in environments with a low reservoir of cold gas.

In \citealt{Torbaniuk:21} (hereinafter \citetalias{Torbaniuk:21}) we presented a first study of the correlation between star formation and AGN activity in the local Universe ($z < 0.33$) using a homogeneous Sloan Digital Sky Survey (SDSS~DR8) optical galaxy sample with robust SFR (in the range $10^{-3}$ to $10^{2}\mathcal{M}_{\odot}$\,year$^{-1}$) and $\mathcal{M}_{\ast}$ estimates (from $10^{6}$ to $10^{12}\mathcal{M}_{\odot}$) in combination with X-ray data from {\it XMM-Netwon} Serendipitous Source Catalogue (3XMM-DR8). This allowed us to estimate the specific BH accretion rate $\lambda_{\mathrm{sBHAR}}$, tracing the level of AGN activity per unit stellar mass of the host galaxy. We found that the local Universe contains a low fraction of efficiently accreting SMBH, while the majority of local SMBH accrete at very low rates. We observed a significant correlation between sBHAR--SFR for almost all stellar masses, but the population of SMBH hosted by quiescent galaxies is accreting at systematically lower levels than star-forming systems. 
This work however suffered from the limited size of the sample and from the low resolution of XMM data, which could affect the estimate of the intrinsic nuclear activity and thus the estimate of the sBHAR, especially for low-luminosity AGN. Thus, in order to confirm the results obtained in \citetalias{Torbaniuk:21}, in the present work we improve the analysis using also X-ray data from the {\it Chandra} Source Catalogue (CSC2.0). Since the {\it Chandra} telescope has the highest resolution among all X-ray telescopes available nowadays, it allows us to better discriminate the nuclear source from the host-galaxy contribution, and test our previous results. Furthermore, the combined 3XMM+CSC2.0 dataset represents the largest serendipitous X-ray survey of the local Universe and will represent a reference sample both for high-z studies as well as for next-generation surveys such as eROSITA \citep{Comparat:22, LiuT:22, Mountrichas:22eR, Aspegren:23, Comparat:23, Mountrichas:23}. 

This paper is organised as follows. In Section~\ref{sec:sdss-data} of this paper we present a short description of the primary catalogue of host galaxy properties from SDSS~DR8. The extraction and correction of new X-ray data from CSC2.0, their cross-calibration with 3XMM-DR8 data, and the completeness of the obtained X-ray sample are discussed in Section~\ref{sec:x-ray-sample}. In Section~\ref{sec:bhar} we present the intrinsic sBHAR probability distribution for star-forming and quiescent galaxies with different stellar masses. In addition, we study X-ray luminosity and $\lambda_{\mathrm{sBHAR}}$ distributions as a function of stellar mass and galaxy properties as well as the correlation between SFR and $\lambda_{\mathrm{sBHAR}}$. We summarise our findings in Section~\ref{sec:concl}. Throughout this paper, we adopt a flat cosmology with $\Omega_{\Lambda} = 0.7$ and ${\rm H}_0 = 70$\;km\;s$^{-1}$ Mpc$^{-1}$.


\section{The SDSS galaxy sample}\label{sec:sdss-data} 

The results in this paper are based on the same initial optical galaxy sample used in \citetalias{Torbaniuk:21}, described fully therein and briefly summarised here. This sample is based on the {\it galSpec} catalogue of galaxy properties\footnote{\url{https://www.sdss.org/dr12/spectro/galaxy\_mpajhu/}} produced by the MPA–JHU group as the subsample from the main galaxy catalogue of the 8th Data Release of the Sloan Digital Sky Survey (SDSS DR8). The stellar masses ($\mathcal{M}_{\ast}$) were obtained through Bayesian fitting of the SDSS {\it ugriz} photometry to a grid of models (see details in \citealt{Kauffmann:03a, Tremonti:04}). The estimates of the SFR were done in two different ways depending on the object classification according to the BPT criteria \citep{Baldwin:81}. The values of SFRs for star-forming galaxies were determined using the H${\alpha}$ emission line luminosity \citep{Brinchmann:04}, while for all other spectral classes (e.g. AGN, composite and unclassified objects) the empirical relation between SFR and the Balmer decrement, D4000, was used (see details in \citealt{Kauffmann:03c}).

The entire {\it galSpec} catalogue provides information for about\,1.5 million galaxies with redshift $z < 0.33$, but in our work we selected only objects with reliable spectroscopic parameters (i.e. with {\tt RELIABLE\,!=\,0}) and redshift estimate (i.e. with {\tt zWarning\,=\,0}). Furthermore, we excluded  duplicates and objects with low-quality SDSS photometry using the basic photometric processing flags\footnote{\url{https://www.sdss.org/dr12/algorithms/photo_flags_recommend/}}. 

To establish the level of AGN activity for different galaxy populations, all galaxies in our sample have been classified as `star-forming' (SFGs) or `quiescent' according to their position on the SFR--$\mathcal{M}_{\ast}$ plain (Fig.\,\ref{fig:sfr-mass-Xray}). Since SFG are found to follow a relatively tight correlation between the current SFR and $\mathcal{M}_{\ast}$, the so-called {\it main sequence} (MS) of star-formation \citep{Noeske:07, Blanton:09, Speagle:14, Renzini:15, Tomczak:16, Santini:17, Pearson:18, Huang:23}, these two classes can be separated using the position of each individual galaxy relative to the evolving MS of SFG. Similarly to the approach used in \citetalias{Torbaniuk:21}, we set the threshold between the two classes 1.3\,dex below the main sequence defined by \citet{Aird:17}, as follows: 
\begin{multline}
\log\mathrm{SFR}_{\mathrm{cut}}(z)\,[\mathcal{M}_{\odot}\mathrm{year}^{-1}] =\\
= -7.6 + 0.76\log\,\mathcal{M}_{\ast}\,/\mathcal{M}_{\odot} + 2.95\log(1+z).
\label{eq:sfg-pass-line}
\end{multline}
{Galaxies that fall above the above threshold (shown by a black band in Fig.\,\ref{fig:sfr-mass-Xray}) were classified as star-forming while those below as quiescent.} Note that the relation in Equation\,\eqref{eq:sfg-pass-line} is redshift-dependent, so the galaxy classification was done considering the redshift of each individual object, { and the range of the thresholds corresponding to galaxies in our redshift range.}

Our final sample consists of 703\,422 galaxies, of which 376\,938 are classified as star-forming (53.6\,per\,cent) and 326\;484 as quiescent galaxies (46.4\,per\,cent), respectively. 

\section{X-ray AGN sample}\label{sec:x-ray-sample}

In order to quantify the AGN activity we need to distinguish the nuclear emission produced by the accretion of the material onto the central SMBH from the stellar emission of the host galaxy, especially in the circumnuclear region. Since the local Universe contains mainly low-to-moderate luminosity AGN, their identification in the optical and IR bands is challenging due to the dominance of the host galaxy emission in these bands. At the same time, the host galaxy's contribution to the total X-ray emission is generally smaller allowing to use of the X-ray as a robust technique for the AGN identification. In \citetalias{Torbaniuk:21} we used 1953 sources with reliable photometry\footnote{We selected detections from the most sensitive PN camera, however, when the data from PN camera were missing, we used those from MOS1 or MOS2 cameras.} and point-like X-ray morphology from the {\it XMM-Newton} Serendipitous Source Catalogue (3XMM-DR8, \citealt{Rosen:16}); their distribution on SFR--$\mathcal{M}_{\ast}$ plane is shown at the top panel of Figure\,\ref{fig:sfr-mass-Xray}. In this work, we present a similar analysis using the X-ray data from {\it Chandra} X-ray Observatory. It has the highest resolution among all X-ray telescopes available nowadays and therefore, allows us to better discriminate the nuclear source from the host-galaxy contribution. To increase the number of studied objects in our sample and improve the statistics of the studied relations we will later combine the data from CSC2.0 and 3XMM-DR8 in the final X-ray sample (see Section\,\ref{sec:cross-calib}). 


\subsection{The {\it Chandra} data set selection}\label{sec:csc-data}

We extracted our data from the {\it Chandra} Source Catalogue\footnote{\url{https://cxc.cfa.harvard.edu/csc/}} (CSC~2.0, \citealt{Evans:10}). Comparing our SDSS galaxy sample (presented in Section\,\ref{sec:sdss-data}) and the {\it Chandra} footprint 
we found that 22\,836 objects from the SDSS sample are falling in the area of the sky observed by the {\it Chandra} X-ray Observatory. 


{In order reduce the X-ray sample size, we used the {\sc CSCview} tool to select all potential X-ray counterparts within a conservative radius of 10\,arcsec, to be sure to include also poorly resolved sources far from the {\it Chandra} optical axis; {\sc CSCview} provided us with coordinates, as well as positional uncertainties (i.e. the position error ellipses\footnote{\url{https://cxc.cfa.harvard.edu/csc/columns/positions.html}}), for 2\,271 sources.
Finally, we refined the cross-match requiring the (optical vs X-ray) source position distance to be less than the sum of the {\it Chandra} positional errors ({\tt err\_ellipse\_r0}, {\tt err\_ellipse\_r1} and {\tt err\_ellipse\_ang}) and the SDSS positional uncertainty of 0.18\,arcsec (i.e. the SDSS fiber position uncertainty, \citealt{Pier:03})\footnote{We used the {\sc TOPCAT} sky ellipse matching algorithm which creates a position ellipse for each source based on the given central position, major and minor radii, and position angle of the ellipse for both databases and compares these elliptical regions on the sky for overlap. See details at \url{http://www.starlink.ac.uk/topcat/} and \citet{Taylor:05:Topcat}.}, reducing the number of objects to 1\,512. The  false match rate for such crossmatch, estimated by randomly shifting our sources by 20\,arcsec and repeating the match, is near 0.4\,per\,cent (i.e. 6 sources for our sample).}

\begin{figure}
\centering
\includegraphics[width=0.95\linewidth]{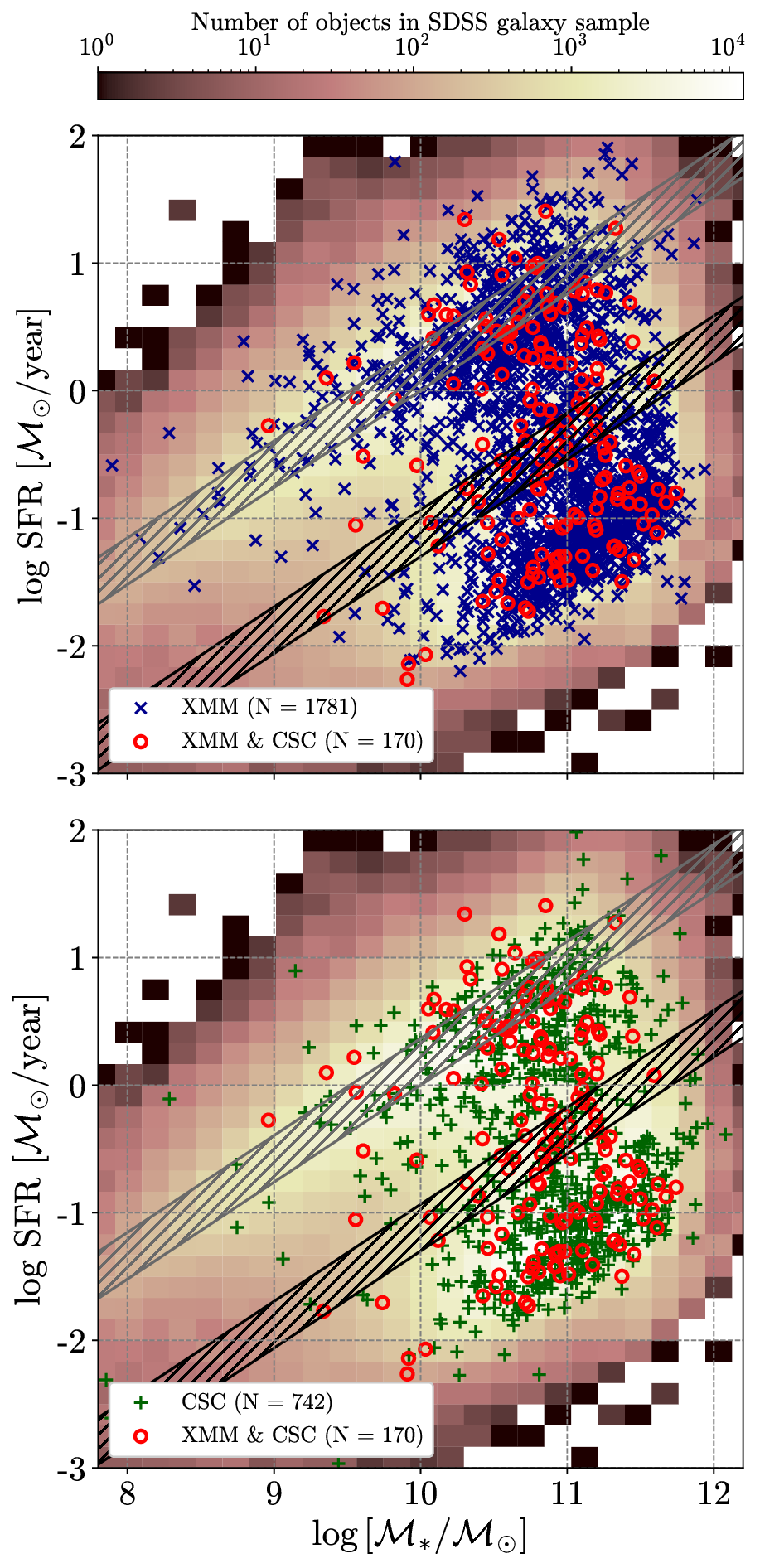}
\vspace*{-1.5ex}
\caption{The distribution of star-formation rate vs. stellar mass for our final SDSS galaxy sample. The grey shaded band shows the main sequence (MS) of star-forming galaxies defined by Eq.\,\eqref{eq:sfg-pass-line}. The black shaded band represents a cut 1.3\,dex below the MS of SFG used for the division of the studied sample into star-forming and quiescent galaxies. Both areas correspond to the redshift interval of our SDSS sample $z = 0.00-0.33$. The individual objects with X-ray detection in the hard band in 3XMM-DR8 (top) and CSC2.0 (bottom) catalogues are shown by blue crosses and green pluses, respectively. Red circles show individual objects that have X-ray detection in both catalogues.}
\label{fig:sfr-mass-Xray}
\end{figure}

To avoid including spatially extended objects (e.g. hot gas regions or galaxy clusters) we only selected objects with zero extension parameter ({\tt extend\_flag == 0})\footnote{However, we did an additional crossmatch of spatially extended sources in the CSC2.0 ({\tt extend\_flag != 0}) with our 3XMM-SDSS sample from \citetalias{Torbaniuk:21} to check how many of them were previously included to our study as `point-like' due to smaller resolution of {\it XMM-Newton} compared to {\it Chandra} observatory (see more details in Section\,\ref{sec:cross-calib}).}. Finally, we choose only the objects with available detection in the hard band (2.0--7.0\,keV) where the AGN emission is likely dominant. As a result, our final CSC-SDSS sample contains 912 objects (454 star-forming and 458 quiescent galaxies); their distribution on SFR--$\mathcal{M}_{\ast}$ plane is shown at the bottom panel of Figure\,\ref{fig:sfr-mass-Xray}.




To compute the rest-frame X-ray luminosities in the hard band (2--7\,keV) we use the aperture-corrected net energy flux in the ACIS hard (2--7\,keV) energy band ({\tt flux\_aper\_h}) available in the CSC Master Source Table\footnote{\url{https://cxc.cfa.harvard.edu/csc/organization.html}}. Following the same step as in \citetalias{Torbaniuk:21} we also applied a K-correction for each value of X-ray luminosities: we assume a photon index $\Gamma = 1.4$ which corresponds to a moderately obscured AGN spectrum with the absorption column density $\log{N_{\rm H}/\mathrm{cm}^{-2}}\simeq 22.5$ (also see \citealt{Tozzi:06, Liu:17}). 


\subsection{Cross-calibration of {\it Chandra} and {\it XMM-Newton} data}\label{sec:cross-calib}

To increase the number of objects in our sample we decided to combine the sample compiled from CSC\,2.0 catalogue (see the previous section) together with the 3XMM-SDSS sample from our \citetalias{Torbaniuk:21} (see sample\,\#4 defined in table\,1) compiled based on 3XMM-Newton Serendipitous Source Catalogue (3XMM-DR8). However, before using the data of these samples together we need to cross-calibrate their fluxes. In order to do so, we used 170 objects from our SDSS sample, which were detected in the hard band both by {\it Chandra} and {\it XMM-Newton} observatories. However, a fraction of nearby galaxies detected as point-like sources in the 3XMM sample may be resolved as spatially extended sources by {\it Chandra} because of its higher resolution compared to {\it XMM-Newton}. We found that an additional 42 point-like sources from our 3XMM-SDSS sample have extended counterparts in the CSC2.0 catalogue. At this stage, we decided to include these 42 sources in our cross-calibration analysis, but they will be excluded from further studies {(i.e. not included in our final CSC-SDSS sample)} because of their X-ray host galaxy contamination; as a result, we have 212 objects for the cross-calibration analysis. 

Since the hard band is represented in slightly different energy ranges in CSC2.0 (2--7\,keV) and in 3XMM (2--12\,keV) we first rescaled all 3XMM fluxes to the energy band used in the CSC2.0. For this, using {\tt WebSpec} tool\footnote{\url{https://heasarc.gsfc.nasa.gov/webspec/webspec.html}} we simulated two spectra within energy ranges of 2--12\,keV and 2--7\,keV assuming a simple power-law model with spectral index $\Gamma \sim 1.4$. As a result, we obtained that the total fluxes obtained for two spectra correlate as $f_{[2-12\,\text{keV}]}/f_{[2-7\,\text{keV}]} = 1.72$. This value was used as a scaling coefficient to transform all 3XMM fluxes available in our sample to the 2--7\,keV energy range. 


Even so, significant discrepancies may still exist between {\it Chandra} and {\it XMM-Newton} due to calibration uncertainties. Therefore, for a proper combination of the data, we used the flux relations obtained by~\citet{Tsujimoto:11} in the cross-calibration analysis of X-ray detections of the pulsar wind nebulae G21.5--0.9. In our work we calibrate fluxes from the three {\it XMM-Newton} pn, MOS1 and MOS2 cameras relative to the flux from {\it Chandra} ACIS camera as $f_{\rm ACIS}/f_{\rm PN} = 1.194$, $f_{\rm ACIS}/f_{\rm MOS1} = 1.108$ and $f_{\rm ACIS}/f_{\rm MOS2} = 1.128$. These calibration coefficients are also in agreement with the results of another cross-calibration analysis based on the sample of galaxy clusters~\citep{Nevalainen:10}.

\begin{figure}
\centering
\includegraphics[width=0.92\linewidth]{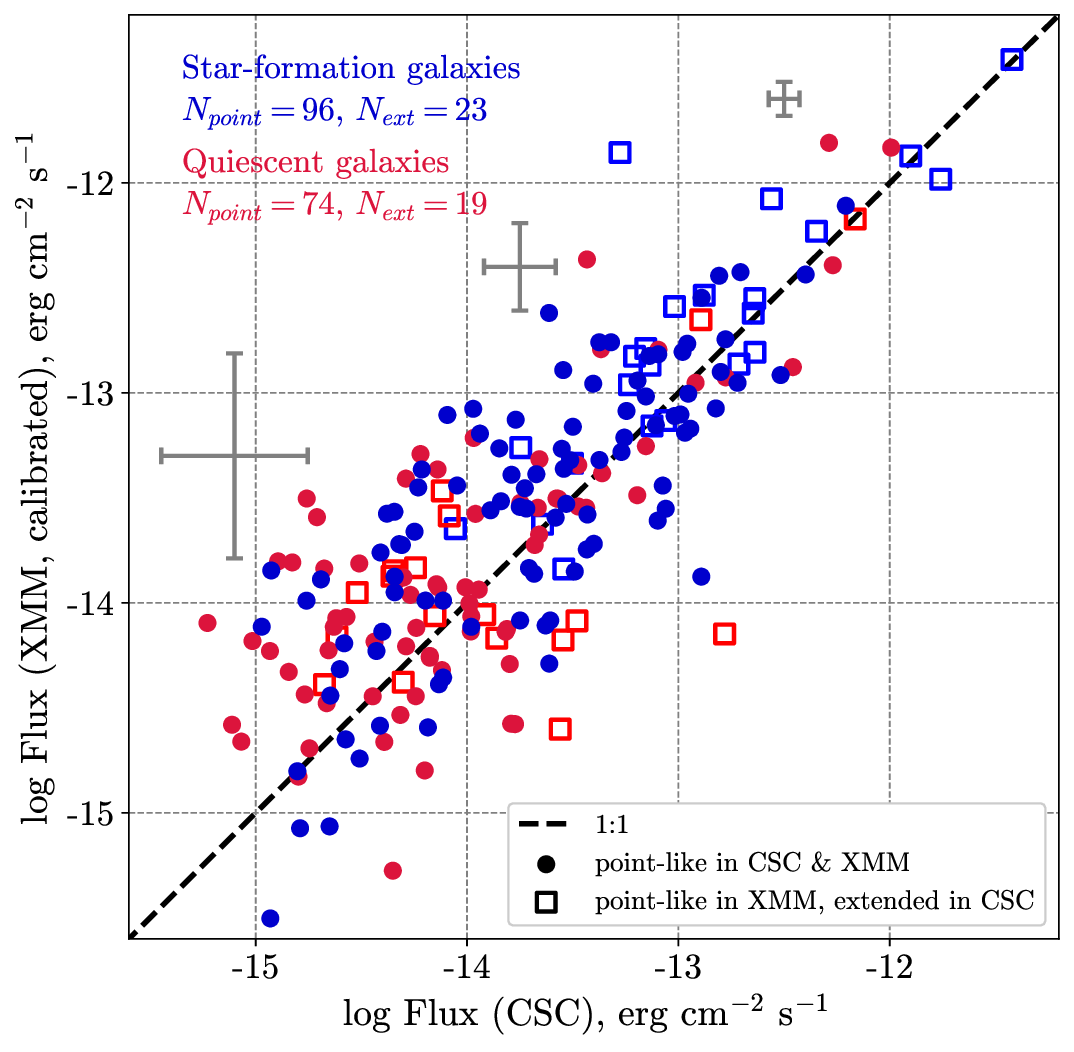}
\vspace*{-1.5ex}
\caption{The relation between the hard-band (2--7\,keV) fluxes in CSC2.0 and 3XMM-DR8 catalogue for 212 galaxies from our SDSS sample (119 star-forming and 93 quiescent galaxies). 170 sources detected as point-like both in the 3XMM and the CSC2.0 are presented by circles, while squares show 42 sources detected as point-like in the 3XMM and as an extended in the CSC2.0. The 3XMM flux is calibrated in accordance with the CSC2.0 flux considering the systematic difference between the two X-ray observatories (see details in the text). }
\label{fig:csc-xmm-calib}
\end{figure}

The comparison between {\it Chandra} fluxes (not-calibrated) and {\it XMM-Newton} fluxes calibrated with the coefficients mentioned above for 212 individual sources is shown in Fig.\,\ref{fig:csc-xmm-calib}. The difference between {\it Chandra} and {\it XMM-Newton} fluxes is $0.4$\,dex on average, exceeding 1\,dex only for a few  extreme cases. The observed scatter is due to the internal uncertainties of each catalogue (the average errors for high, medium, and low flux values are presented by grey markers in Fig.\,\ref{fig:csc-xmm-calib}), combined with the uncertainties in the flux conversion factors and the scatter introduced by variability of  individual nuclear sources. In conclusion, we assume that the two calibration steps described above are sufficient for our study, as the residual systematics will not affect our final results (see Section\,\ref{sec:bhar}). 

Based on the obtained results, we applied the same calibration corrections to the 3XMM fluxes of 1\,741 from 1\,953 objects detected only in the 3XMM-SDSS sample. The rest of the objects have been detected also by {\it Chandra}, so for 170 `true' point-like objects (both in the 3XMM-DR8 and CSC2.0) we further use the hard band flux from the CSC2.0.
As a result, the combined sample contains 2\,653 sources. {Hereinafter, we refer to the sources with only 3XMM detections as 3XMM sources and those with CSC2.0 detections as CSC2.0 sources, even though some of them also have 3XMM detections.}


\subsection{Contribution of the host galaxy to the total X-ray emission}\label{sec:corr_sfg_etg}

To assess the level of AGN activity from the X-ray emission we need first to determine the contribution of the host galaxy to the total X-ray luminosity. The {\it Chandra} telescope has a higher resolution than {\it XMM-Newton} and is able to more easily separate the nuclear emission from the host-galaxy contribution, the faintest objects or the galaxies at greater distances (i.e. smaller angular size of the object). In principle, we should estimate the fractional contribution of the host galaxy within the region where the X-ray flux is measured; however, since the scaling relations we use are computed for the entire galaxy and the X-ray flux extraction radii are different for each source based on the position within the telescope field of view, we followed a simpler approach already used for 3XMM data in \citetalias{Torbaniuk:21} estimating an upper limit to the corrections assuming that the contribution is due to the entire galaxy. 

Since our sample contains both star-forming and quiescent galaxies (see definition in Section\,\ref{sec:sdss-data}) we need to consider the contributions of different galaxy components to its total X-ray emission. For {\it star-forming} galaxies we calculated the expected X-ray luminosities ($L_\mathrm{X,host}$) due to low and high-mass X-ray binaries (LMXB and HMXB) based on the scaling relation between $L_{\mathrm{X}}$ and SFR, stellar masses $\mathcal{M}_{\ast}$ and redshift $z$ of galaxies from \citet{Lehmer:16}. On the contrary, {\it quiescent} galaxies {are dominated by hot gas, low-mass X-ray binaries (LMXB), and some emission from coronally active binaries (ABs) and cataclysmic variables (CVs)}; see \citet{Fabbiano:06, Boroson:11, Kim:13, Civano:14, Jones:14}. To eliminate the contribution of LMXB and AB+CV we used a relation of $L_\mathrm{X}$ and galaxy luminosity in the $K_{S}$-band from \citet{Boroson:11}, while for the hot gas, we chose to use $L_\mathrm{X}-L_\mathrm{K_{S}}$ relation defined in \citet{Civano:14}. The galaxy luminosity in the $K_{S}$-band was calculated based on $K_{S}$ magnitudes from the 2MASS\footnote{\url{https://old.ipac.caltech.edu/2mass/}} Point and Extended Source Catalogues \citep{Skrutskie:06}. The detailed description of $K_{S}$-band luminosity calculation and all applied relations are presented in \citetalias{Torbaniuk:21}.

\begin{figure*}
\includegraphics[width=0.99\linewidth]{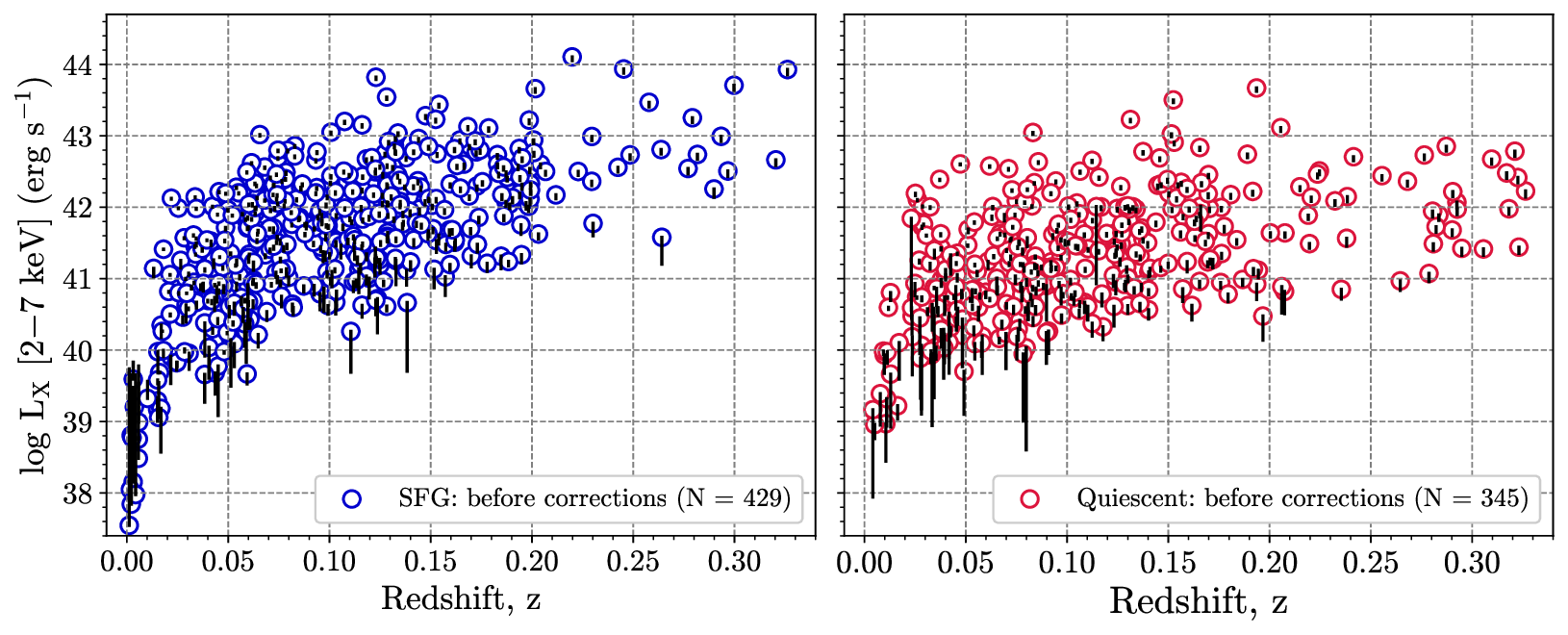}
 \vspace*{-1.5ex}
\caption{The X-ray luminosity vs redshift distribution of 774 objects with positive residual X-ray luminosity from CSC-SDSS sample. The {observed (uncorrected)} L$_\mathrm{X}$ values for SFGs and quiescent galaxies are presented as blue and red circles, respectively. The change in L$_\mathrm{X}$ {due to the subtraction of the predicted host galaxy contribution (i.e. L$_\mathrm{X,host}$, see description in the text)} for each object is shown by a solid line.}
\label{fig:Lum-corr}
\end{figure*}

{In this work we label an object as an AGN if its observed `total' X-ray luminosity exceeds the value of the predicted for its host galaxy derived as explained above. Thus to isolate the AGN contribution, we subtracted the host-galaxy contribution from the total X-ray luminosity of the source. As a result, we identify 2\,223 AGN with positive residual X-ray luminosity after the correction 
1\,449 of which are 3XMM-only sources and 774 are CSC2.0 sources (the latter including 144 sources in common with 3XMM sources).
The total X-ray luminosity (and the value of correction for the host galaxy contribution) vs redshift distribution in the hard band, for star-forming and quiescent galaxies,  
are shown in Fig.\,\ref{fig:Lum-corr}. Since the same distribution for 3XMM sources has been already shown in \citetalias{Torbaniuk:21}, in this paper we present distributions only for objects with available detection in the CSC2.0 catalogue. 
It is worth mentioning that both the observed X-ray luminosity and the scaling relations used to evaluate the host galaxy contribution have their own uncertainties, and therefore sources with relatively low luminosities can be falsely excluded (or included) from our sample using the criterion of the positive residual X-ray luminosity. We found that changing the threshold by the uncertainty in the scaling relations affects only 1.5\,per\,cent of SFG and 5.3\,per\,cent of quiescent galaxies in our sample, while 6.7\,per\,cent of all sources have luminosities consistent with the threshold within the photometric errors. Thus the fraction of objects potentially affected by these uncertainties is relatively small (11.4\,per\,cent in total); furthermore, the effect due to the scaling relation uncertainty is systematic in nature, while the photometric uncertainties are statistical, so the results presented in Section\,\ref{sec:bhar} will be scarcely affected.}

\subsection{The stellar mass completeness of CSC+XMM sample}\label{sec:completeness}

\begin{figure}
\centering
\includegraphics[width=0.99\linewidth]{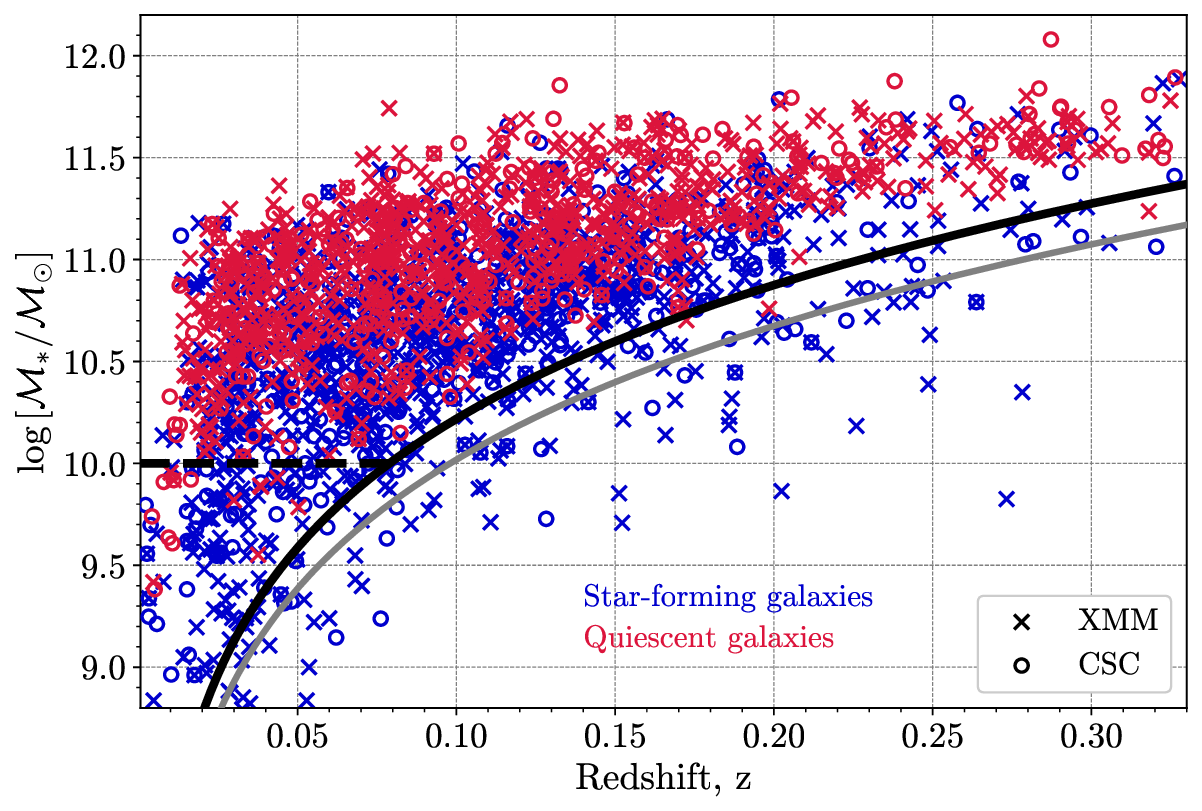}
\vspace*{-1.5ex}
\caption{The host galaxy stellar mass vs redshift distribution of combined CSC+XMM sample, where sources with X-ray detection in CSC2.0 and 3XMM presented by crosses and circles, respectively. Red and blue colors show quiescent and star-forming galaxy populations, respectively. The grey solid curve shows the redshift-dependent mass limit for a galaxy with a maximally old stellar population at limiting SDSS magnitude $r = 17.5$\,mag. The horizontal dashed and solid black lines show the limit of $10^{10}\mathcal{M}_{\odot}$ and 0.2\,dex above the derived SDSS mass limit, respectively, used to remove the part of the sample with significantly lower fraction of X-ray detected quiescent galaxies with respect to SFG (see the text for details).}
\label{fig:mass-completeness}
\end{figure}

Our SDSS galaxy sample is magnitude-limited to a Galactic extinction-corrected Petrosian magnitude of $r=17.5$ (see SDSS~DR7 Target selection page\footnote{\url{https://classic.sdss.org/dr7/products/general/target_quality.html}}). As a result, star-forming and quiescent galaxies have different $\mathcal{M}_{\ast}$ limits because of the different mass-to-light ratios of their stellar populations. This introduces incompleteness as quiescent galaxies of a given stellar mass drop out of the sample at lower redshifts compared to SFG of the same stellar mass. This source of bias is well visible plotting stellar mass as a function of redshift separately for star-forming and quiescent galaxies (see Figure\,\ref{fig:mass-completeness}). To minimise this effect of bias we used the same approach as proposed in~\citet{Georgakakis:14} applying a redshift-dependent mass limit which corresponds to a maximally old (i.e. maximal mass-to-light ratio) galaxy. We used the \citealt{Bruzual:03} model of the mass-to-light ratio evolution considering Kroupa~IMF \citep{Kroupa:01} and for each redshift, we calculated the value of stellar mass that corresponds to an observed magnitude of $r = 17.5$~mag (see solid grey line in Fig.\,\ref{fig:mass-completeness}). Above this limit, the galaxy sample should not be affected by the incompleteness of the survey because all galaxies have a mass-to-light ratio smaller than that of the maximally old stellar population model. However, this limit refers to the entire SDSS galaxy sample and does not represent the `real' limit for our subsample of X-ray detected sources. In order to allow a fair comparison, we chose to exclude regions in Fig.\,\ref{fig:mass-completeness} where the fraction of X-ray detected quiescent galaxies is significantly lower than the fraction of SFG. Thus, we limit our study to higher mass values $>10^{10}\mathcal{M}_{\odot}$ and 0.2\,dex above the derived SDSS limit (see dashed and solid black lines in Fig.\,\ref{fig:mass-completeness}).

As a result, the final XMM+CSC sample contains 1\,938 objects (967 star-forming and 971 quiescent galaxies).


\section{The specific Black Hole accretion rate}\label{sec:bhar}

Following the approach presented in \citetalias{Torbaniuk:21} we calculate the specific Black Hole accretion rate ($\lambda_{\mathrm{sBHAR}}$), the rate of accretion onto the central SMBH scaled relative to the stellar mass of the host galaxy. We followed the definition from \citet{Bongiorno:12, Bongiorno:16, Aird:18}: 
\begin{equation}
    \lambda_{\mathrm{sBHAR}} = \frac{k_{\mathrm{bol}}\,{L}_{\mathrm{X,hard}\,}[\mathrm{erg\,s}^{-1}]}{1.3\cdot10^{38}\;[\mathrm{erg\,s}^{-1}]\times0.002\;{\mathcal{M}_{\ast}}\,/{\mathcal{M}_{\odot}}},
    \label{eq:spec-bhar}
\end{equation}
where $k_{\mathrm{bol}}$ is a bolometric correction factor for the hard band, ${L}_{\mathrm{X,hard}}$ is the 2--7\,keV X-ray luminosity. Although the bolometric correction factor is dependent on the luminosity \citep{Marconi:04, Lusso:10, Lusso:12, Bongiorno:16, Duras:20}, here we adopted an average bolometric correction of $k_{\mathrm{bol}} = 25$ since our sample is not probing the range of high bolometric luminosities, where $k_{\mathrm{bol}}$ increases significantly with respect to $25$, and thus the other systematics discussed below dominate the final uncertainty. We assumed that the  Black Hole mass scales with the host galaxy stellar mass as $\mathcal{M}_{\mathrm{BH}} = 0.002\,\mathcal{M}_{\ast}/\mathcal{M}_{\odot}$ as in \citet{Haring:04}. The additional scale factors are defined from the request that $\lambda_{\mathrm{sBHAR}}\approx\lambda_{\mathrm{Edd}}$, where the Eddington ratio $\lambda_{\mathrm{Edd}} \propto \mathrm{L}_{\mathrm{X}}/\mathcal{M}_{\mathrm{BH}}$.


\subsection{The sBHAR distribution as a function of stellar mass}\label{sec:bhar-distr}

To study the distribution of $\lambda_{\rm sBHAR}$ in the local Universe we measure  
$p(\log \lambda_{\rm sBHAR} | \mathcal{M_{\ast}})$, which represents the probability density function (PDF) that a galaxy with a certain stellar mass hosts a SMBH accreting with a given $\lambda_{\rm sBHAR}$. In what follows, $p(\log \lambda_{\rm sBHAR} | \mathcal{M_{\ast}})$ was calculated for the full galaxy sample as well as separately for star-forming and quiescent galaxies, allowing to evaluate the difference in AGN activity for galaxy populations characterised by different morphology, star-formation history, gas content and transportation processes. 

\subsubsection{Methodology} 
\begin{figure}
\centering
\includegraphics[width=0.99\linewidth]{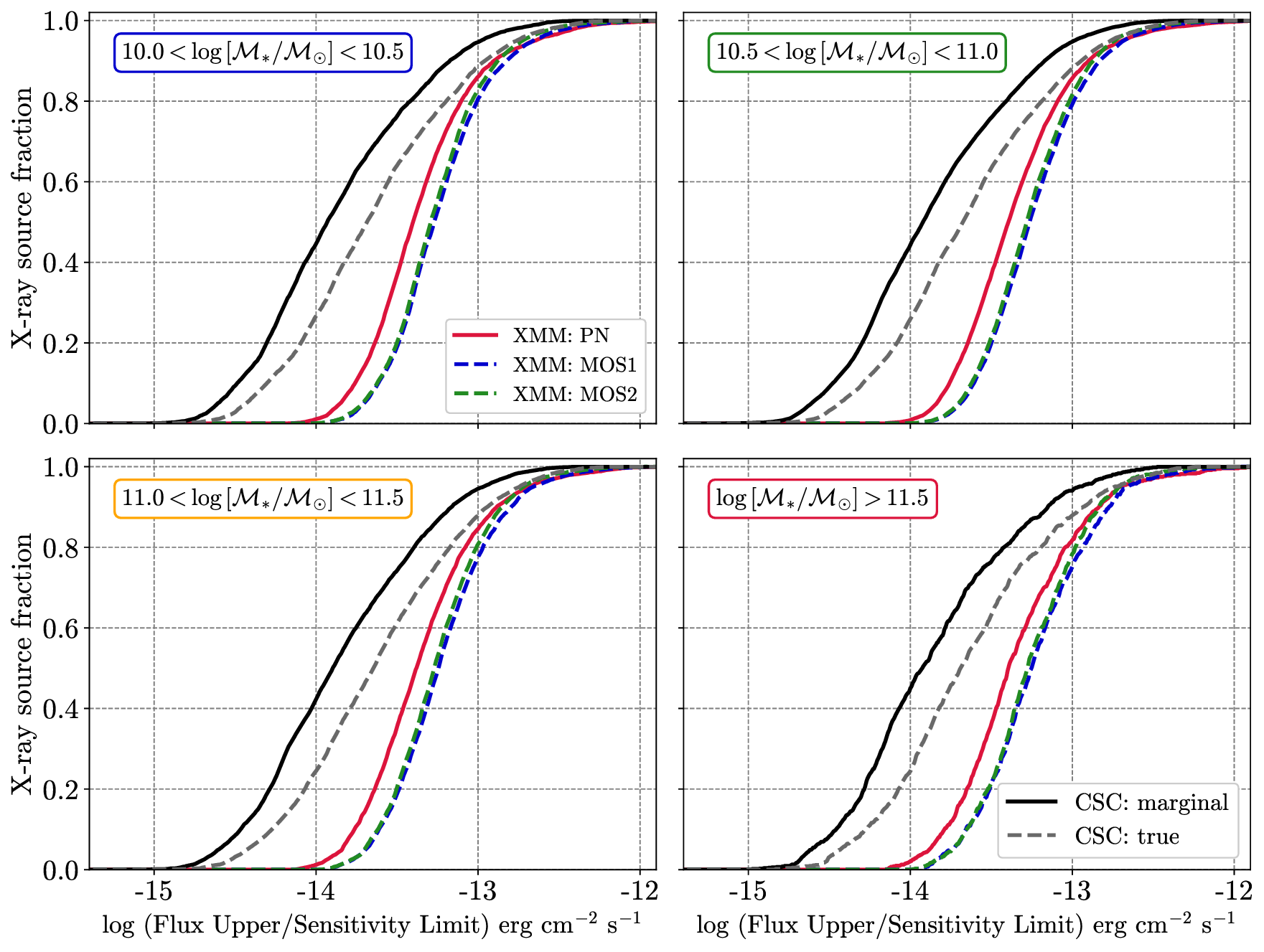}
\vspace*{-1.5ex}
\caption{The cumulative histogram of flux upper-limits in the hard band (2--12\,keV) for three XMM cameras (pn, MOS1 and MOS2 cameras by red, blue and green color, respectively) from the XMM FLIX service and of flux sensitivity-limits in the hard band (2--7\,keV) for {\sc MARGINAL} (black solid) or {\sc TRUE} (black dashed) detections in the CSC2.0 for four stellar mass ranges.}
\label{fig:cumul-csc}
\end{figure}

\begin{figure*}
\centering
\includegraphics[width=0.99\linewidth]{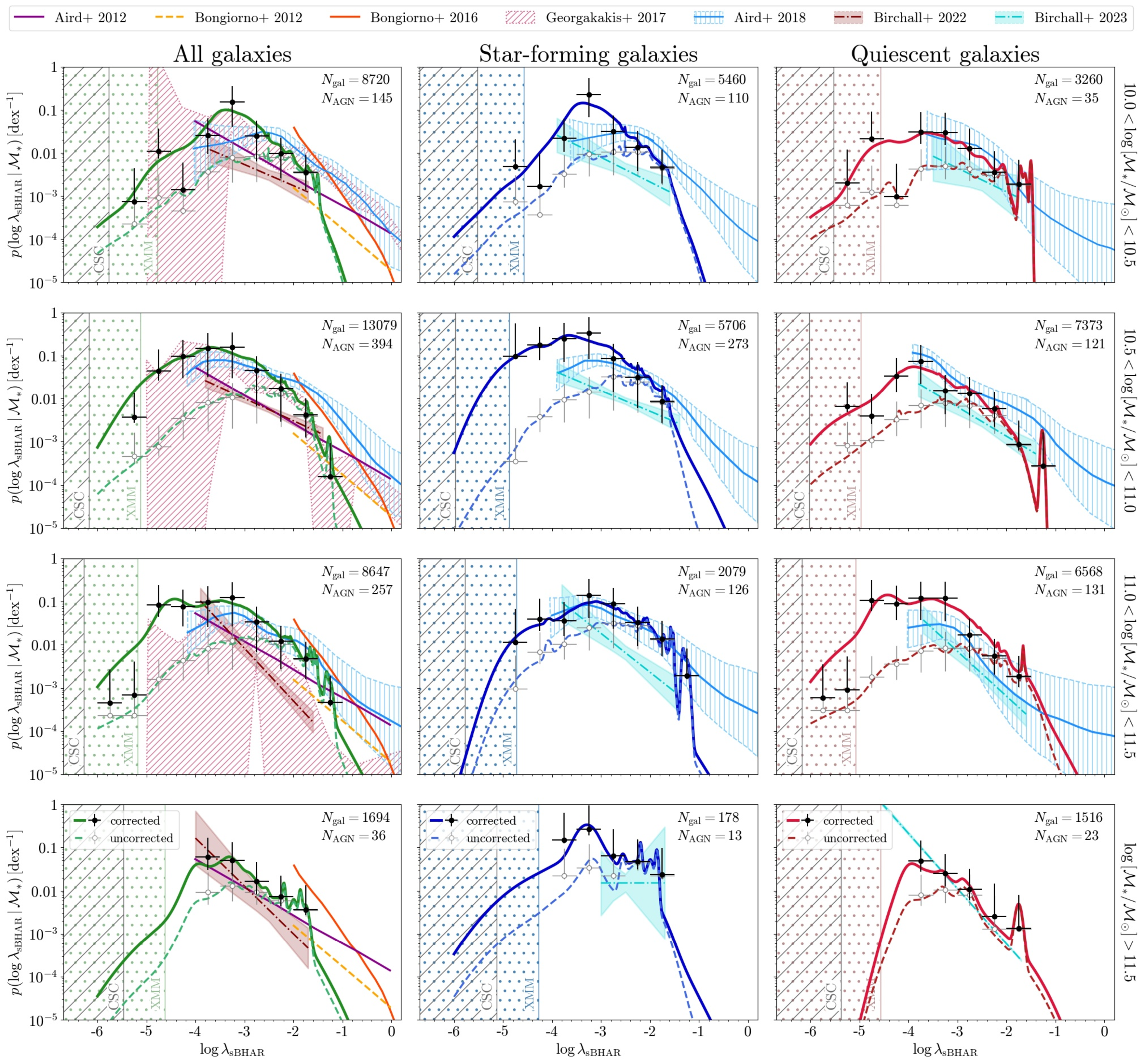}
\vspace*{-1.5ex}
\caption{The probability distribution of specific Black Hole accretion rates, $\log\lambda_{\rm sBHAR}$, as a function of the stellar mass of the host galaxy (increasing from top to bottom) for all (both star-forming and quiescent, left column), star-forming (center) and quiescent (right) galaxies. The black and grey points showed the same probability distribution obtained from the observed distributions of $\log\lambda_{\rm sBHAR}$, while the solid and dashed lines represent the probability distribution obtained as a sum of Gaussian distributions for each individual AGN in the studied sample (see detailed description in the text). The total number of galaxies falling inside X-ray footprints and the total number of AGN with hard X-ray detection in each stellar mass range are given in the legend of each panel. The sBHAR detection limits for marginal detection in CSC2.0 catalogue and the most sensitive detection in the PN camera in the 3XMM catalogue are shown by shaded grey (CSC) and dotted color (XMM) areas, respectively. The power-law fits of sBHAR distributions estimated by \citealt{Aird:12} ($z = 0.6$, $9.5 < \log\,[\mathcal{M}_{\ast}/\mathcal{M}_{\odot}] < 12.0$) and \citealt{Bongiorno:12, Bongiorno:16} ($0.3 < z < 0.8$, $8.0 < \log\,[\mathcal{M}_{\ast}/\mathcal{M}_{\odot}] < 12.0$) are shown with violet solid, yellow dashed and orange solid lines, respectively (lines are identical for all stellar mass panels). The sBHAR probability distributions for three stellar mass ranges ($\log\,[\mathcal{M}_{\ast}/\mathcal{M}_{\odot}] = $ 10.0--10.5, 10.5--11.0 and 11.0--11.5) and $z = $~0.0--0.5 obtained by \citet{Georgakakis:17} are presented by pink shaded areas which corresponds to 90\,per\,cent confidence intervals. The light blue solid line and corresponding shaded area (90\,per\,cent confidence interval) show the sBHAR distributions obtained by \citet{Aird:18} within $0.1 < z < 0.5$ for three stellar mass ranges for star-forming, quiescent and all galaxies. {The power-law fits (and their 1$\sigma$ uncertainty) of sBHAR distributions of \citet{Bi22, Bi23} are shown by the dash-dotted lines (and corresponding dotted areas) for all galaxies (brown, \citealt{Bi22}) and separately for SFG and quiescent (cyan, \citealt{Bi23}) within four stellar mass ranges and $z < 0.3$.} }
\label{fig:histo-bhar-csc-xmm}
\end{figure*}

As we mentioned before, while our X-ray sample contains sources above a certain detection likelihood in the full band, to compute the intrinsic nuclear accretion rate distribution we use sources reliably detected in the hard band, so that we can compute a proper completeness correction.
In order to consider the fraction of missed sources as a function of flux, we need to account for the sensitivity variations of the X-ray observations covering our SDSS galaxy sample across the sky (due to different detector efficiency, exposure time, off-axis angle, etc). 
Since our CSC+XMM sample is a mixture of two X-ray catalogues, the sensitivity corrections are  estimated separately for each X-ray sample.  For sources with X-ray detection only in the 3XMM-DR8 catalogue, only 405 of 1\,257 sources have a reliable likelihood detection {\sc DET\_ML > 6} ($3\sigma$) in the hard band.
We collected the values of the survey flux sensitivity limit (from XMM FLIX, \citealt{Carrera:07}) at the position of each source in our original optical sample falling in the 3XMM footprint. The cumulative curves in four stellar mass ranges are shown in Figure\,\ref{fig:cumul-csc}. In essence, the three curves describe the likelihood of detecting the X-ray counterpart of our galaxies at each flux level in each of the three XMM cameras (pn, MOS1, and MOS2).
These cumulative values were applied as statistical weights to the number of sources used to compute $p(\log\,\lambda_{\rm sBHAR}|\mathcal{M}_{\ast})$; see later Figure\,\ref{fig:histo-bhar-csc-xmm}.

On the other hand, the source detection process in the CSC2.0 catalogue does not provide the source detection likelihood directly, but the classification as {\sc FALSE}, {\sc MARGINAL} or {\sc TRUE} from the analysis of stacked images\footnote{The {\sc MARGINAL} and {\sc TRUE} source detection likelihood thresholds are detected from simulations and correspond to false source rates of $\sim$1 and $\sim$0.1 false sources per stack, respectively. More detailed description of this process can be found in the section `Limiting sensitivity and Sky coverage' on the webpage of the statistical properties of the CSC2.0 catalogue: \url{https://cxc.harvard.edu/csc/char.html}}. Therefore, we used the 427 sources (out of 681 with X-ray detection in the CSC2.0), with hard fluxes higher than the corresponding {\sc MARGINAL} or {\sc TRUE} flux sensitivity limits provided by {\tt CSCview} service. The two additional CSC cumulative curves in Figure\,\ref{fig:cumul-csc}  represent the likelihood of detecting the {\sc MARGINAL} or {\sc TRUE} X-ray counterpart of our galaxies with a given flux in the hard band in the CSC2.0. These cumulative curves were applied as statistical weights to the CSC2.0 detected sources, as done for XMM sources above. 

The binned corrected distribution of sBHAR in our hard X-ray galaxy sample, in the $-6 < \log\,\lambda_{\rm sBHAR} < 0$ range, is presented in Figure\,\ref{fig:histo-bhar-csc-xmm} for the entire galaxy population (left panel), and separately for star-forming and quiescent galaxies (central and right panels), in four stellar mass ranges. The errors for each probability point were calculated using the confidence limits equation from \citet{Gehrels:86}.

We note that, in comparison with the more advanced analysis presented in other works using deep surveys (e.g. \citealt{Bongiorno:16, Georgakakis:17, Aird:18}), we are using only X-ray fluxes and upper limits. This prevents us from using more complex approaches like, e.g., Bayesian modelling, which requires the availability of individual photons (e.g. source and background counts) instead of the archival data products. As an alternative, we estimated a continuous probability distribution function assuming that the likelihood of observing the certain value of $\log\,\lambda_{\rm sBHAR}$ in a given galaxy can be described by a normal  distribution centered on the best $\log\lambda_{\rm sBHAR}$ estimate and with width derived from flux error of each AGN using the same host galaxy parameters ($z$ and $\mathcal{M}_{\ast}$).
The total probability distribution was then estimated as the sum of the individual PDFs for all objects in the corresponding stellar mass range ($N_{\rm AGN}$) normalised by the total number of galaxies (within the same mass range) falling inside the X-ray footprint ($N_{\rm gal}$). To correct the effect of the variable sensitivity across, each individual PDF was normalized using the statistical weight from the cumulative curves in Figure\,\ref{fig:cumul-csc}. As a result, both uncorrected and corrected $p(\log\,\lambda_{\rm sBHAR}|\mathcal{M}_{\ast})$ are presented Figure\,\ref{fig:histo-bhar-csc-xmm} by dashed and solid lines, respectively. 

Finally, to quantify the detection limit of our data, we estimated the minimum sBHAR $\log\lambda_{\rm sBHAR, min}$, that can either be detected with {\it XMM-Newton} or corresponding to a `marginal' detection in CSC2.0. In order to do this, for each individual galaxy falling inside the 3XMM and CSC2.0 footprint ($N_{\rm gal}$) we use the minimum flux sensitivity from the cumulative curves in Figure\,\ref{fig:cumul-csc} and converted it to the lowest detectable sBHAR through Eq.\,\eqref{eq:spec-bhar}, using the proper $z$ and $\mathcal{M}_{\ast}$. From the cumulative distribution of $\log\lambda_{\rm sBHAR, min}$ in each stellar mass range we define our detection limit as the sBHAR value for which there is a probability of detecting at least one AGN in our sample. The obtained sBHAR detection limits for marginal detection in the CSC2.0 and the most sensitive detection by PN camera in the 3XMM are shown in Figure\,\ref{fig:histo-bhar-csc-xmm} by shaded grey and dotted color areas, respectively.

\subsubsection{The analysis of the local sBHAR distribution and its comparison with the literature} 

The completeness-corrected $p(\log\,\lambda_{\mathrm{sBHAR}}|\mathcal{M}_{\ast})$ distribution in Fig.\,\ref{fig:histo-bhar-csc-xmm} has an approximately power-law shape with flattening (or even turnover) toward low accretion rates for all stellar mass ranges indicating the prevalence of low-efficiency accretion in the local Universe. This trend is broadly consistent with the studies of the $\lambda_{\mathrm{sBHAR}}$ probability functions {presented in \citet{Bi22, Bi23}}, derived via non-parametric models as in \citet{Aird:12, Bongiorno:12, Georgakakis:17}, by adopting analytic models for the sBHAR distribution convolved with the galaxy mass function as in \citet{Bongiorno:16}, or using the Bayesian mixture modelling approach as in \citet{Aird:18}. 

\begin{table*}
\centering
 \caption{The values of AGN fraction, $f(\log\lambda_{\rm sBHAR} > -5.0)$ and $f(\log\lambda_{\rm sBHAR} > -2.0)$, and the average specific accretion rate, $\log\,\langle\lambda_{\rm sBHAR}\rangle$, for four stellar mass ranges for all, star-forming and quiescent galaxies. The estimates are derived by integrating $p(\log\lambda_{\rm sBHAR}|\mathcal{M}_{\ast})$ distributions presented in Figure\,\ref{fig:bhar-sfr-csc-xmm} (see also the description in the text).}
 \label{tab:fract-avg-agn}
 \begin{tabular}{ccccccccccc}
  \hline
\multirow{2}{*}{\#} & \multirow{2}{*}{Stellar mass range} & \multicolumn{3}{c}{$f(\log\lambda_{\rm sBHAR} > -5.0)$ [\%]} & \multicolumn{3}{c}{$f(\log\lambda_{\rm sBHAR} > -2.0)$ [\%]} & \multicolumn{3}{c}{$\log\,\langle\lambda_{\rm sBHAR}\rangle$}\\[2pt]
\cline{3-5}\cline{6-8}\cline{9-11}\\[-2ex]
& & All & Star-forming & Quiescent & All & Star-forming & Quiescent & All & Star-forming & Quiescent \\
\hline
1 & $10.0 < \log\,[\mathcal{M}_{\ast}/\mathcal{M}_{\odot}] < 10.5$ & 11.3 & 15.1 & 5.0 & 0.2 & 0.3 & 0.1 & $-3.82$ & $-3.69$ & $-4.21$ \\
2 & $10.5 < \log\,[\mathcal{M}_{\ast}/\mathcal{M}_{\odot}] < 11.0$ & 24.8 & 47.3 & 7.3 & 0.4 & 0.7 & 0.07 & $-3.62$ & $-3.32$ & $-4.24$\\
3 & $11.0 < \log\,[\mathcal{M}_{\ast}/\mathcal{M}_{\odot}] < 11.5$ & 20.7 & 15.9 & 22.3 & 0.3 & 0.9 & 0.2 & $-3.68$ & $-3.40$ & $-3.83$ \\
4 & $\log\,[\mathcal{M}_{\ast}/\mathcal{M}_{\odot}] > 11.5$ & 6.9 & 27.2 & 4.5 & 0.2 & 1.1 & 0.09 & $-4.06$ & $-3.34$ & $-4.37$ \\
  \hline
 \end{tabular}
\end{table*}

We find that the $p(\log\,\lambda_{\mathrm{sBHAR}}|\mathcal{M}_{\ast})$ distributions for all galaxies and stellar mass ranges cover a wide range of the BH accretion rates pointing that the variability of the AGN activity happens in shorter timescales compared to the long-term host galaxy processes. This AGN variability can be the result of the stochastic nature of the processes responsible for the gas transportation to the nuclear region of galaxies, as well as of AGN and stellar feedback processes that are able to heat and/or remove the gas from the nuclear region, thus preventing its accretion onto the SMBH.
The peak of the derived sBHAR distribution in the local Universe occurs at low BH accretion rates $-4 \leq \log\,\lambda_{\mathrm{sBHAR}} \leq -3$, which tends to be offset for quiescent galaxies $\log\,\lambda_{\mathrm{sBHAR}} \approx -4$ relative to SFG ($\log\,\lambda_{\mathrm{sBHAR}} \approx -3$). This finding agrees with results obtained by \citet{Georgakakis:14, Aird:18, Bi22, Bi23} for low redshift samples. {The lower normalisation of the sBHAR distributions of \citet{Bi22, Bi23} with respect to ours can be possibly due to differences in the completeness correction for the X-ray sensitivity variations (i.e. difference in the X-ray correction energy band) used in these works with respect to us.}  
At the same time, Figure\,\ref{fig:histo-bhar-csc-xmm} shows that the probability of a galaxy to host the SMBH accreting at relatively high sBHAR (i.e. $\log\,\lambda_{\rm sBHAR} > -2$) is smaller in the local Universe compared to the studies at high redshifts \citep{Bongiorno:12, Bongiorno:16, Georgakakis:17, Aird:18} likely due to a smaller amount of the gas available for AGN feeding or the rarity of large-scale events (e.g. galaxy interaction and mergers) able to trigger intensive gas supply to the central regions necessary to fuel high accretion rate AGN. However, it needs to be noted that the lack of sources with  $\log\,\lambda_{\mathrm{sBHAR}} \geq -1$ in our sBHAR distributions is also due to the fact that our optical sample by definition excludes bright AGNs (Seyfert~1 and quasars), where the AGN continuum dominates the host galaxy emission. On the contrary, at low BH accretion rates $\log\,\lambda_{\rm sBHAR} \leq -2$ the shape of the distribution begins to flatten and possibly turnover for $\log\,\lambda_{\rm sBHAR} \leq -4$ similarly to those distributions obtained by \citet{Georgakakis:17, Aird:18}. Such behavior of the sBHAR distribution at lower BH accretion rates most likely reflects the natural lower limit of AGN activity, i.e. the minimal fuelling level necessary to trigger radiatively efficient AGN observed in X-ray. However, probing the turnover of the sBHAR distribution is quite challenging firstly because of its proximity to the sensitivity limits of current X-ray telescopes, but also due to the difficulty of separating the nuclear emission from the host galaxy, which becomes dominant in X-ray at such luminosities. 

Finally, based on the $p(\log\,\lambda_{\mathrm{sBHAR}}|\mathcal{M}_{\ast})$ distributions we can derive the AGN fraction $f(\log\lambda_{\rm sBHAR})$, which is fully accounted for the varying sensitivity of the X-ray observations across the sky, representing the fraction of galaxies in the local Universe that contain a central black hole that is accreting above a certain limit in $\log\lambda_{\rm sBHAR}$. In order to do this, we integrate our estimates of $p(\log\,\lambda_{\mathrm{sBHAR}}|\mathcal{M}_{\ast})$ down to two limits of $\log\lambda_{\rm sBHAR} = -5.0$ and $-2.0$. These two sBHAR limits were chosen so as to evaluate the fraction of the `entire' X-ray selected AGN population (with $\log\lambda_{\rm sBHAR} > -5.0$, i.e. down to the CSC sensitivity limit) and the fraction of galaxies hosting AGN with moderate-to-high accretion rates, i.e. black holes are growing above $\sim 1$ per\,cent of their Eddington limit ($\log\lambda_{\rm sBHAR} > -2.0$). The estimated AGN fractions for all, star-forming and quiescent galaxy populations in four stellar mass ranges are presented in Table\,\ref{tab:fract-avg-agn}. As we see, the fraction of AGN with $\log\lambda_{\rm sBHAR} > -5.0$ lies in the range from 7 to 24~per\,cent, while the fraction of so-called `classical' AGN with moderate-to-high accretion rate reaches only 0.4~per\,cent, which supports the fact of the dominance of low-efficiency accretion in the local Universe. Moreover, we found that star-forming galaxies show higher AGN fractions in almost all stellar mass ranges (15\;--\;47~per\,cent) relative to quiescent galaxies (ranging from 5 to 22~per\,cent), which may indicate the difference in accretion modes and/or mechanisms responsible for AGN fuelling for different galaxy populations. At the same time, AGN fractions do not show a strong tendency to increase with stellar mass both for SFG and quiescent galaxies. {These findings are also in agreement with low redshift results in \citet{Aird:18,Bi22,Bi23}}. To quantify the average accretion rate in the local Universe we calculate the value of $\langle\lambda_{\rm sBHAR}\rangle$ within a given galaxy sample and stellar mass range by integrating $\lambda_{\rm sBHAR}\times p(\log\,\lambda_{\mathrm{sBHAR}}|\mathcal{M}_{\ast})$ distribution over the entire studied range of sBHAR. The analysis of the derived values of $\log\langle\lambda_{\rm sBHAR}\rangle$ in Table\,\ref{tab:fract-avg-agn} showed that SFG with different stellar mass seem to maintain accretion at the same rate, while quiescent galaxies tend to have lower values of $\log\,\langle\lambda_{\rm sBHAR}\rangle$ with a weak tendency to increase with stellar mass.


\subsection{The X-ray luminosity and sBHAR correlation with stellar mass and star-formation rate}\label{sec:bhar-sfr-chandra}

\subsubsection{Stellar mass}

\begin{figure*}
\centering
\includegraphics[width=0.48\linewidth]{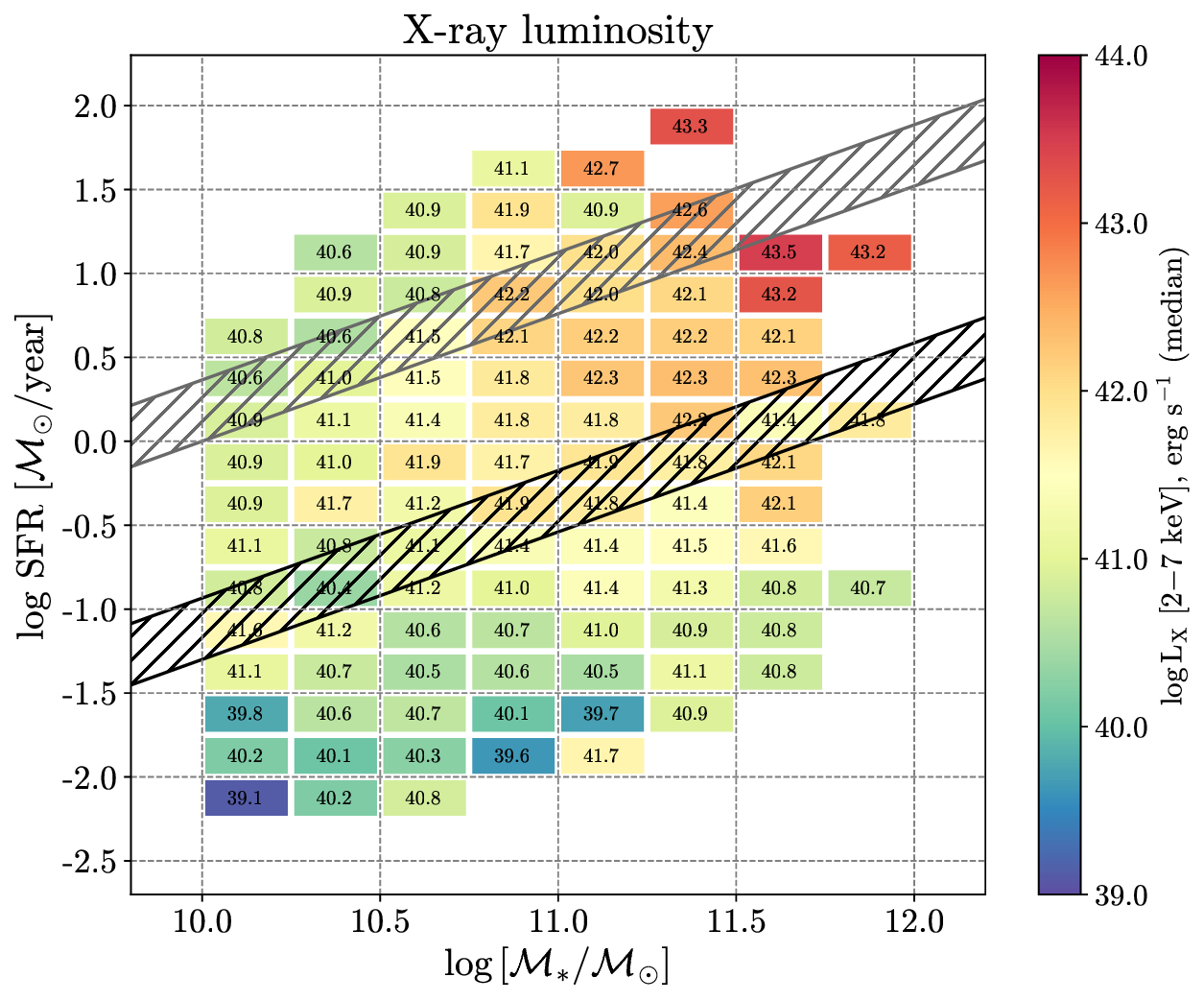}
\includegraphics[width=0.48\linewidth]{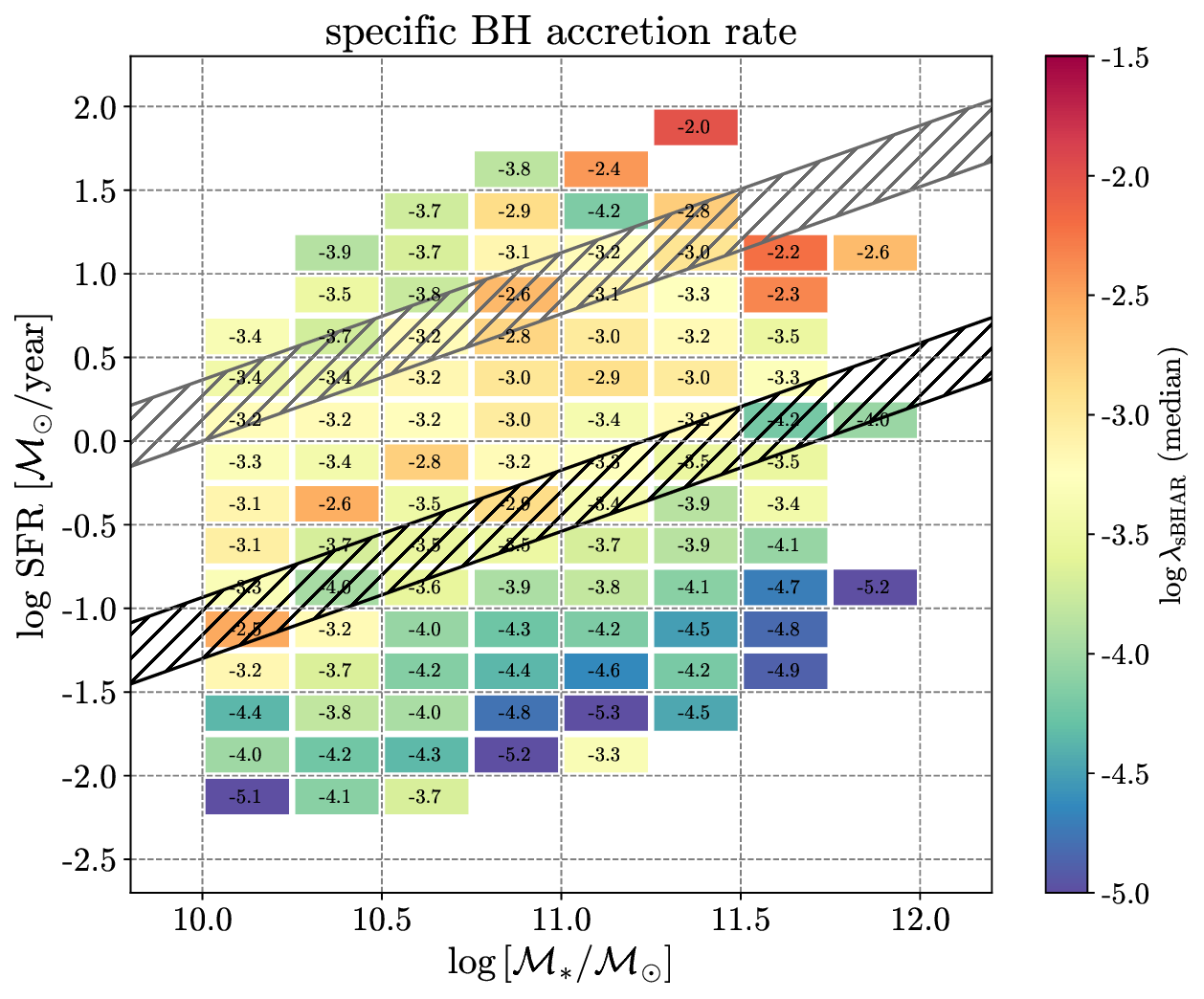}
\vspace*{-1.5ex}
\caption[The distribution of X-ray luminosity and the specific BH accretion rate on SFR--$\mathcal{M}_{\ast}$ plane for {\it Chandra} AGN sample]{The distribution of X-ray luminosity ({\it left}) and the specific BH accretion rate $\lambda_{\mathrm{sBHAR}}$ ({\it right}) on SFR--$\mathcal{M}_{\ast}$ plane for {\it Chandra} AGN sample. The actual median value of $\lambda_{\mathrm{sBHAR}}$ (X-ray luminosity) for each bin of SFR and $\mathcal{M}_{\ast}$ is written inside the square. The black and grey shaded areas are the same as in Figure\,\ref{fig:sfr-mass-Xray}. The number of points in both diagrams ranges from 104 in the central part to 2-3 in the edges. }
\label{fig:median-bhar-sfr-m-csc-xmm}
\end{figure*}

In the previous section, we see that AGN in the local Universe show a broad range of accretion rates. Therefore, to analyse the dependence between AGN activity and the properties of the host galaxy (i.e. the total stellar mass and SFR), following the same steps as in \citetalias{Torbaniuk:21}, we divided our CSC+XMM sample in bins of SFR and $\mathcal{M}_{\ast}$ (with binwidth of 0.25\,dex) and calculated the median $\lambda_{\mathrm{sBHAR}}$ and $L_{\mathrm{X}}$ in each bin. The resulting distribution of $\lambda_{\mathrm{sBHAR}}$ (and $L_{\mathrm{X}}$) on the SFR--$\mathcal{M}_{\ast}$ diagram is shown in Figure\,\ref{fig:median-bhar-sfr-m-csc-xmm}. 

The figure shows that the median value of ${L}_{\rm X}$ increases with $\mathcal{M}_{\ast}$ both for star-forming and quiescent galaxies (left panel of Fig.\,\ref{fig:median-bhar-sfr-m-csc-xmm}). This trend is consistent with our previous results in \citetalias{Torbaniuk:21} and also those found in \citet{Mullaney:12b, Delvecchio:15, Heinis:16, Carraro:20, Stemo:20} showing that more massive galaxies have the tendency to host AGN with higher X-ray luminosity than galaxies with smaller stellar masses. On the contrary, the relation between median $\log\,\lambda_{\rm sBHAR}$ and stellar mass (see the right panel of Fig.\,\ref{fig:median-bhar-sfr-m-csc-xmm}) is not so straightforward. For instance, star-forming galaxies show similar values of the median $\log\,\lambda_{\rm sBHAR}$ for all stellar mass ranges, while the median sBHAR for quiescent galaxies seems to decrease with $\mathcal{M}_{\ast}$. Similar weak correlation between sBHAR and $\mathcal{M}_{\ast}$ was also found in Table\,\ref{tab:fract-avg-agn} and \citetalias{Torbaniuk:21}. At the same time, a number of studies showed that BH accretion rate correlates positively with stellar mass at different redshift \citep{Rosario:13, Xu:15, Yang:18, Carraro:20}; however, this relation seems to become weaker toward the local Universe \citep{Yang:18} due to increasing number of massive quiescent galaxies with less abundant cold gas fuelling the less luminous AGN \citep{Rosario:13, Aird:17}. In addition, the weakening of the sBHAR-$\mathcal{M}_{\ast}$ relation toward lower redshifts may be a product of strong AGN evolution with redshift, whereby AGN feedback in the form of wind produced by high-luminous AGN may expel the cold gas from the host galaxy and thus reduce BH accretion (i.e. the self-regulation of the SMBH growth in massive galaxies). It should be also mentioned, that previous sBHAR-$\mathcal{M}_{\ast}$ relations studied in the literature usually focus on highly accreting AGN (e.g. quasars; especially in high redshift studies), which are missing in our sample. 

\subsubsection{Star-formation rate}

Analysing two different galaxy populations in Figure\,\ref{fig:median-bhar-sfr-m-csc-xmm} we found that quiescent galaxies possess a smaller median sBHAR (and X-ray luminosity) with respect to star-forming galaxies at fixed $\mathcal{M}_{\ast}$ (identically to the trend obtained from $p(\log\lambda_{\rm sBHAR}|\mathcal{M}_{\ast})$ in the previous Section; see Table\,\ref{tab:fract-avg-agn}). A similar difference of $\log\,\lambda_{\mathrm{sBHAR}}$ for the two different galaxy populations was also presented in \citet{Delvecchio:15, Rodighiero:15, Aird:18} and \citetalias{Torbaniuk:21}, and can be explained by a scenario where both star-formation and AGN activity are triggered by fuelling from a common cold gas reservoir \citep{Alexander:12}. We need to note that the separation into star-forming and quiescent galaxies in this work has been done assuming a linear cut 1.3\,dex below the MS of SFG (see Section\,\ref{sec:sdss-data}), but in reality galaxy populations in the local Universe usually show mixed properties (e.g. late-type spiral galaxies with quenched SF vs star-forming galaxies with elliptical early-type morphology in \citealt{Paspaliaris:23}).

\begin{figure*}
\centering
\includegraphics[width=0.95\linewidth]{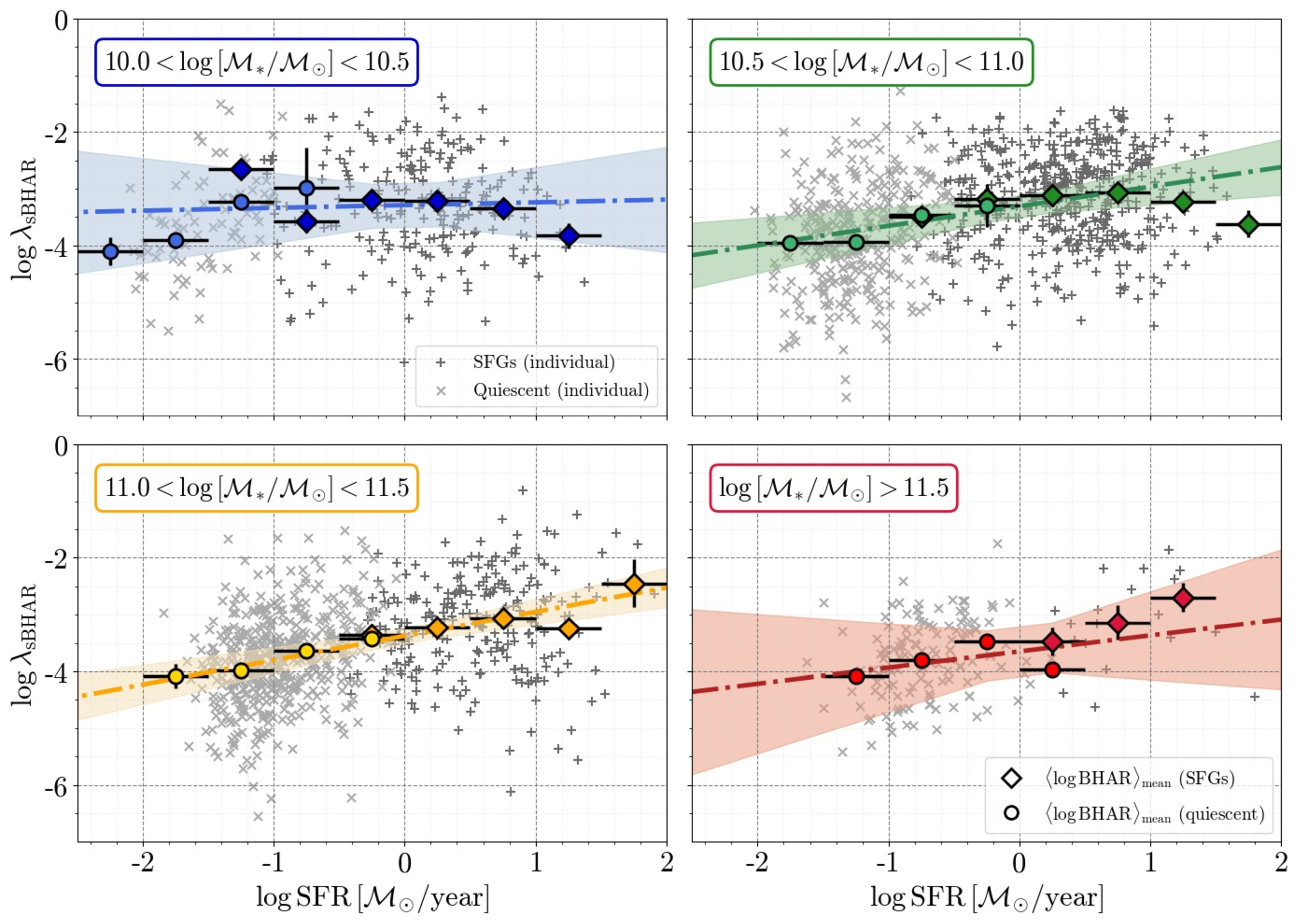}
\vspace*{-1.5ex}
\caption{The jackknife mean value of sBHAR versus SFR for star-forming (diamond) and quiescent galaxies (circles) for four stellar masses ranges. The individual objects from our X-ray AGN sample are represented by grey crosses (SFGs) and pluses (quiescent). The errorbars were calculated as a variance of the jackknife mean. The dash-dotted line shows the least-square linear best fit with 95\,per\,cent confidence interval. The best fitting and goodness-of-fit parameters are presented in Table\,\ref{tab:wls-fit-all-z1-z2}.}
\label{fig:bhar-sfr-csc-xmm}
\end{figure*}

To probe the relation between SFR and sBHAR in more detail we calculated the mean $\langle\lambda_{\mathrm{sBHAR}}\rangle$ in 10 bins of SFR in the range $-2.5 < \log \mathrm{SFR} < 2.0$, where the uncertainty of $\langle\log\,\lambda_{\mathrm{sBHAR}}\rangle$ was computed using jackknife resampling. The $\langle\log\,\lambda_{\mathrm{sBHAR}}\rangle$--$\log\,\mathrm{SFR}$ correlation is presented in four stellar mass ranges in Figure\,\ref{fig:bhar-sfr-csc-xmm}. To all derived values of $\langle\lambda_{\mathrm{sBHAR}}\rangle$ we applied the regression analysis and fitted the sBHAR-SFR correlation using the linear approximation and least-squares regression model. The best fit parameters are listed in Table\,\ref{tab:wls-fit-all}. 
Figure\,\ref{fig:bhar-sfr-csc-xmm} confirms a  trend of increasing sBHAR with SFR from quiescent to star-forming galaxies for all stellar mass ranges. However, the linear regression analysis confirms that the $\langle\log\,\lambda_{\rm sBHAR}\rangle$ correlates with SFR at $>95$\,per\,cent confidence ($P$-value $<0.05$) only for two intermediate stellar mass intervals (the intervals \#2\,and\,3 on Table\,\ref{tab:wls-fit-all}). At the same time, at the two lowest $\mathcal{M}_{\ast}$ ranges (\#1 and \#2) sBHAR seems to increase with SFR only for quiescent galaxies, while for star-forming galaxies $\langle\lambda_{\mathrm{sBHAR}}\rangle$-SFR relation flattens and possibly shows a drop for $\log\,{\rm SFR} \gtrsim 1$. The comparison of the sBHAR-SFR relation with those available in the literature for low and high redshift AGN samples is presented in the following Section\,\ref{sec:disc-comparison}.

\begin{table*}
\centering
 \caption{The best fit parameters obtained from a linear relation between $\langle\log\,\lambda_{\mathrm{sBHAR}}\rangle$ and $\log\,\mathrm{SFR}$ for four stellar mass ranges for the CSC+XMM sample (see Figure\,\ref{fig:bhar-sfr-csc-xmm}). The slope, intercept with their standard errors, and all statistics parameters ($F$-statistic, $P$ value, and $R^{2}$) were found from the least-square linear regression. In this work, we consider the confidence level as $P$-value $<0.05$ (stellar mass ranges satisfying this criterion are marked in blue). $N$ is the number of points in each stellar mass bin.}
 \label{tab:wls-fit-all}
 \begin{tabular}{cccccccc}
  \hline
\# & Stellar mass range & slope & intercept & {\it F}-statistic & $P$ value ({\it F}-stat) & $R^{2}$ & $N$ \\
\hline
1 & $10.0 < \log\,[\mathcal{M}_{\ast}/\mathcal{M}_{\odot}] < 10.5$    & $0.05\pm0.13$ & $-3.29\pm0.14$ & 0.14 & 0.7231 & 0.02 & 10 \\[1pt] 
2 & $10.5 < \log\,[\mathcal{M}_{\ast}/\mathcal{M}_{\odot}] < 11.0$    & $0.34\pm0.07$ & $-3.31\pm0.07$ & 22.44 & {\color{blue} 0.0015} & 0.74 & 10 \\[1pt] 
3 & $11.0 < \log\,[\mathcal{M}_{\ast}/\mathcal{M}_{\odot}] < 11.5$    & $0.43\pm0.05$ & $-3.38\pm0.05$ & 72.58 & {\color{blue} $6.1\cdot10^{-5}$} & 0.91 & 9 \\[1pt] 
4 & $\log\,[\mathcal{M}_{\ast}/\mathcal{M}_{\odot}] > 11.5$           & $0.28\pm0.18$ & $-3.65\pm0.13$ & 2.61 & 0.1670 & 0.34 & 7 \\[1pt] 
  \hline
 \end{tabular}
\end{table*}

\subsubsection{The influence of selection effects on the sBHAR-SFR relation}

In interpreting the flattening of $\langle\log\,\lambda_{\rm sBHAR}\rangle$--$\log\,{\rm SFR}$ relation for star-forming galaxies with relatively low stellar masses (see $10.0 < \log\,[\mathcal{M}_{\ast}/\mathcal{M}_{\odot}] < 10.5$ and $10.5 < \log\,[\mathcal{M}_{\ast}/\mathcal{M}_{\odot}] < 11.0$ ranges in Fig.\,\ref{fig:bhar-sfr-csc-xmm}) we need to be aware of the selection effects affecting our sample. As it can be seen in Fig.\,\ref{fig:mass-completeness} the lowest stellar mass range (i.e. $10.0 < \log\,[\mathcal{M}_{\ast}/\mathcal{M}_{\odot}] < 10.5$) contains only nearby objects (i.e. redshift $z < 0.1$) and misses objects at higher redshifts. At the same time, the highest stellar mass range (i.e. $\log\,[\mathcal{M}_{\ast}/\mathcal{M}_{\odot}] > 11.5$) shows a dearth of  massive objects at $z < 0.15$. Since both sBHAR and SFR present a rapid decline toward $z \sim 0$ \citep{Delvecchio:14, Madau:14, Aird:15, DSilva:23}, the observed difference could reflect differences in the sample average evolutionary stage.

\begin{figure}
\centering
\includegraphics[width=0.99\linewidth]{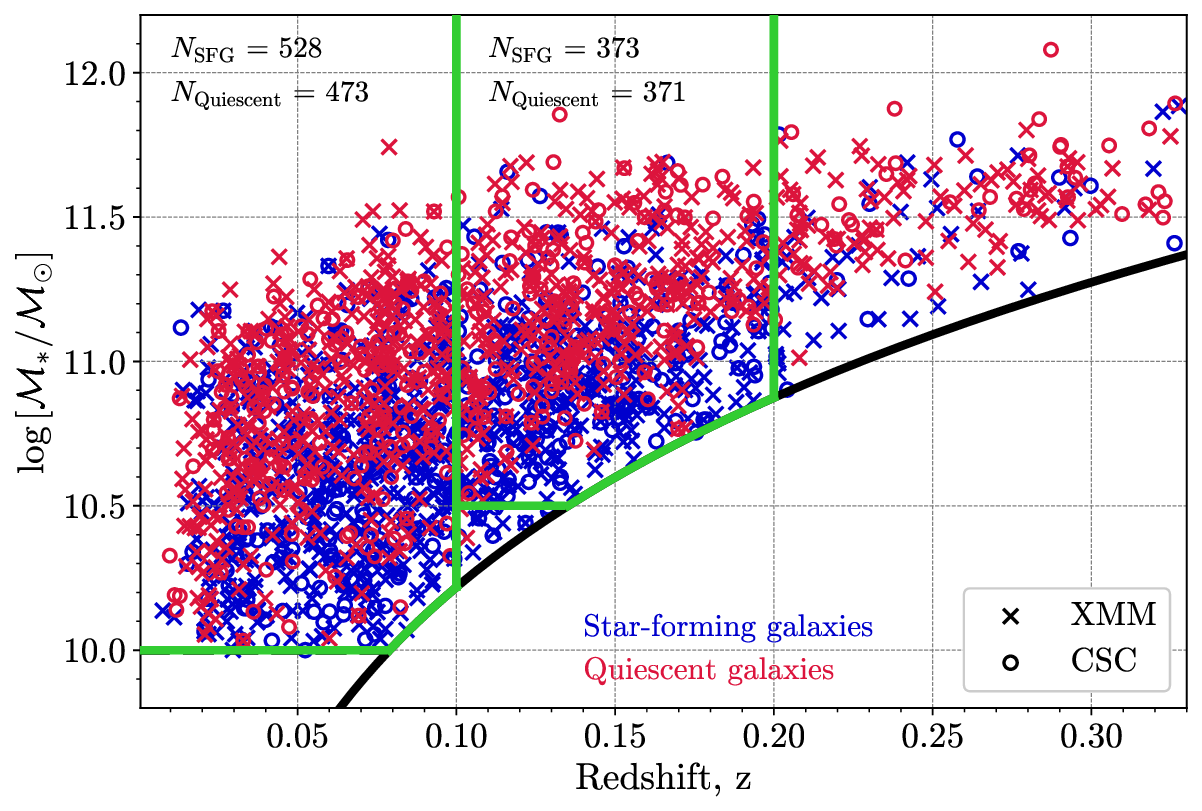}
\vspace*{-1.5ex}
\caption{The same host galaxy stellar mass vs redshift distribution as in Figure\,\ref{fig:mass-completeness}. The green solid lines show two selected subsamples in two redshift intervals $z < 0.1$ and $0.1 < z < 0.2$. The horizontal lines show the mass limits $10^{10}$ and $10^{10.5} \mathcal{M}_{\odot}$ for these two subsamples.} 
\label{fig:mass-compl-z}
\end{figure}

To check the influence of the selection effect mentioned above on the shape of $\langle\log\,\lambda_{\rm sBHAR}\rangle$--$\log\,{\rm SFR}$ relation we use two redshift-limited subsamples as shown in Fig.\,\ref{fig:mass-compl-z}: the first subsample is limited to $z < 0.1$ with stellar mass  $\log\,[\mathcal{M}_{\ast}/\mathcal{M}_{\odot}] > 10$, while the second one contains objects within $0.1 < z < 0.2$ and $\log\,[\mathcal{M}_{\ast}/\mathcal{M}_{\odot}] > 10.5$. As a result, we obtained 1\,001 objects in the $z < 0.1$ subsample (528 SFGs and 473 quiescent galaxies) and 744 objects in the second $0.1 < z < 0.2$ subsample (373 SFGs and 371 quiescent). 

\begin{figure*}
\centering
\includegraphics[width=0.85\linewidth]{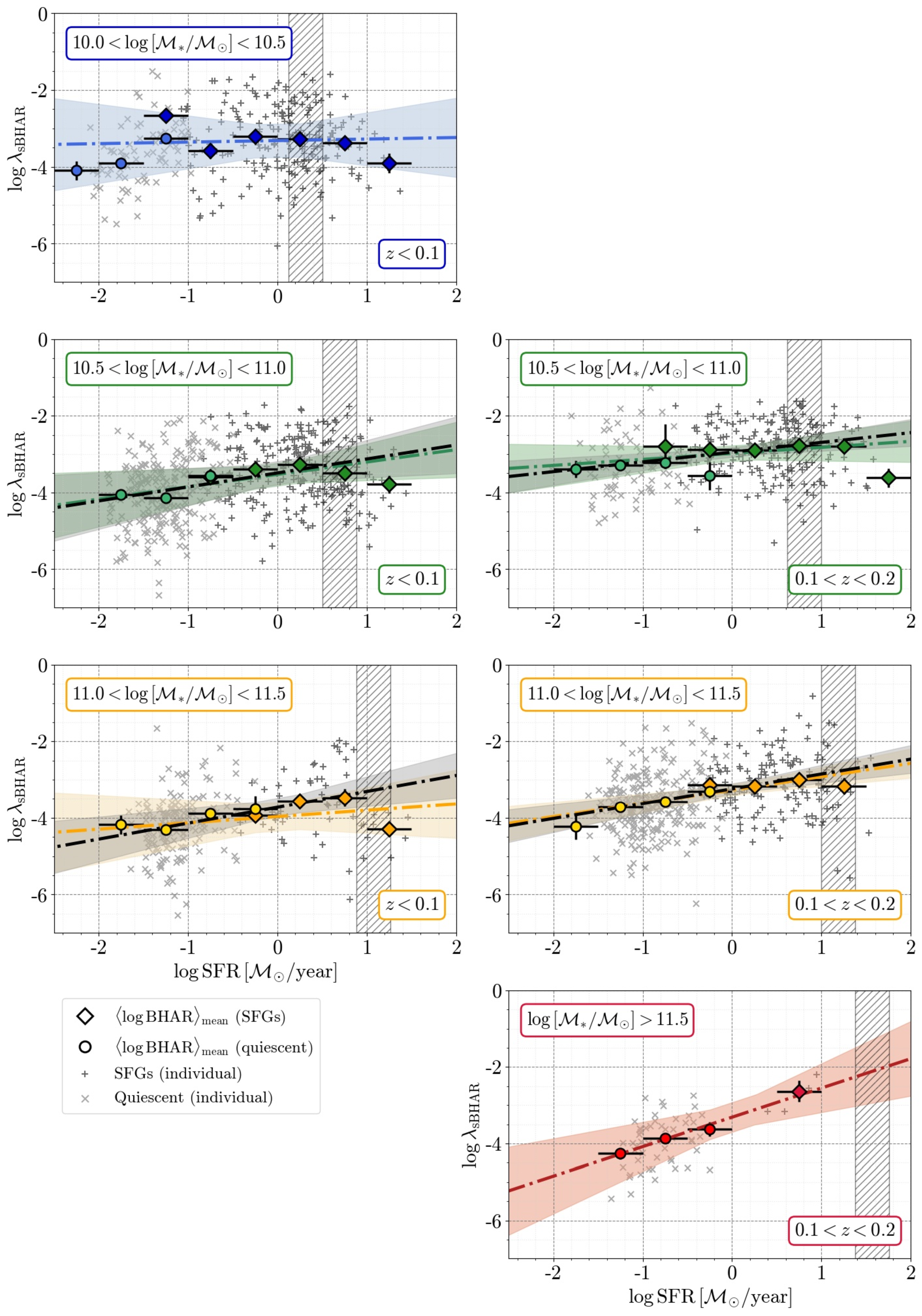}
\caption{The same jackknife mean value of sBHAR vs SFR as in Fig.\,\ref{fig:bhar-sfr-csc-xmm} for star-forming (diamond) and quiescent galaxies (circles) in the redshift intervals $z < 0.1$ ({\it left}) and $0.1 < z < 0.2$ ({\it right}). The black dash-dotted line shows the least-square linear best fit with 95\,per\,cent confidence interval (grey) considering only points with $\log\,{\rm SFR} < 1.0$. The grey shaded area represents the position of the MS of SFG, defined 1.3\,dex above the cut used for the star-forming and quiescent galaxies separation (see definition in Eq.\,\eqref{eq:sfg-pass-line} and the text of Section\,\ref{sec:sdss-data}). To plot the MS area for each separate panel we used the extreme values of $\log\,[\mathcal{M}_{\ast}/\mathcal{M}_{\odot}]$ for each stellar mass range and the maximum value of $z$ for the corresponding redshift subsample. }
\label{fig:bhar-sfr-z1-z2}
\end{figure*}

Following a similar approach as in the previous section we computed $\langle\lambda_{\rm sBHAR}\rangle$ in SFR bins separately for $z < 0.1$ and $0.1 < z < 0.2$ redshift subsamples. The $\langle\lambda_{\rm sBHAR}\rangle$--$\log\,{\rm SFR}$ relation and the best fitting parameters of its linear approximation are shown in Fig.\,\ref{fig:bhar-sfr-z1-z2} and Table\,\ref{tab:wls-fit-all-z1-z2}. A statistically significant $\langle\log\,\lambda_{\rm sBHAR}\rangle$--$\log\,{\rm SFR}$ correlation at $>95$\,per\,cent confidence level ($P$ value $< 0.05$) was confirmed only for one stellar mass range in $z < 0.1$ (\#2: $10.5 < \log\,[\mathcal{M}_{\ast}/\mathcal{M}_{\odot}] < 11.0$) and for two stellar mass ranges in $0.1 < z < 0.2$ (\#3 and 4: $11.0 < \log\,[\mathcal{M}_{\ast}/\mathcal{M}_{\odot}] < 11.5$ and $\log\,[\mathcal{M}_{\ast}/\mathcal{M}_{\odot}] > 11.5$, respectively). However, it should be noted that the stellar mass range \#4 for $0.1 < z < 0.2$ contains only 5 star-forming galaxies with $\log\,{\rm SFR} > 0$ and therefore, the best-fit may not reflect the intrinsic slope of the sBHAR-SFR relation in this stellar mass range. Compared to the $\langle\lambda_{\rm sBHAR}\rangle$--$\log\,{\rm SFR}$ relation for the entire sample (see Fig.\,\ref{fig:bhar-sfr-csc-xmm}) the results for two redshift-limited subsamples show the presence of the sBHAR drop at higher SFR for almost all stellar mass ranges (see in Fig.\,\ref{fig:bhar-sfr-z1-z2}). This suggests that the lack of the drop observed in Figure\,\ref{fig:bhar-sfr-csc-xmm} is related to the fact that at high masses we are probing higher redshifts where the $\langle\lambda_{\rm sBHAR}\rangle$--$\log\,{\rm SFR}$ relation can be different (see Sec.\,\ref{sec:disc-comparison}). Moreover, repeating the regression analysis for the stellar mass ranges \#2 and \#3 considering points only with $\log\,{\rm SFR} < 1$ we found that $\langle\log\,\lambda_{\rm sBHAR}\rangle$ positively correlates with $\log\,{\rm SFR}$ at $>95$\,per\,cent confidence level for both redshift intervals and both stellar mass ranges (see best-fit parameters in square brackets in Table\,\ref{tab:wls-fit-all-z1-z2} and the grey area in Fig.\,\ref{fig:bhar-sfr-z1-z2}).

\begin{table*}
\centering
 \caption{The same as in Table\,\ref{tab:wls-fit-all}, but for two redshift intervals $z < 0.1$ and $0.1 < z < 0.2$ presented in Fig.\,\ref{fig:bhar-sfr-z1-z2}. The values in square brackets correspond to the best fit parameters obtained from a linear $\langle\log\,\lambda_{\rm sBHAR}\rangle$--$\log\,{\rm SFR}$ relation considering only points with $\log\,{\rm SFR} < 1.0$.}
 \label{tab:wls-fit-all-z1-z2}
 \begin{tabular}{ccccccccc}
  \hline
$z$ interval & \# & Stellar mass range & slope & intercept & {\it F}-statistic & $P$ value ({\it F}-stat) & $R^{2}$ & $N$ \\
\hline
\multirow{6}{*}{$z < 0.1$} & 1 & $10.0 < \log\,[\mathcal{M}_{\ast}/\mathcal{M}_{\odot}] < 10.5$    & $0.04\pm0.14$ & $-3.32\pm0.15$ & 0.08 & 0.7901 & 0.01 & 9 \\[1pt] 
& 2 & $10.5 < \log\,[\mathcal{M}_{\ast}/\mathcal{M}_{\odot}] < 11.0$    & $0.32\pm0.10$ & $-3.53\pm0.10$ & 10.71 & {\color{blue} 0.0170} & 0.64 & 8 \\[1pt] 
&  &  & [$0.36\pm0.10$] & [$-3.49\pm0.09$] & [14.30] & [{\color{blue} 0.0129}] & [0.74] & [7] \\[1pt] 
& 3 & $11.0 < \log\,[\mathcal{M}_{\ast}/\mathcal{M}_{\odot}] < 11.5$    & $0.16\pm0.12$ & $-3.97\pm0.12$ & 1.87 & 0.2211 & 0.24 & 8 \\[1pt] 
&  &  & [$0.42\pm0.08$] & [$-3.72\pm0.08$] & [30.15] & [{\color{blue} 0.0027}] & [0.86] & [7] \\[1pt] 
& 4 & $\log\,[\mathcal{M}_{\ast}/\mathcal{M}_{\odot}] > 11.5$           & -- & -- & -- & -- & -- & -- \\[1pt] 
\hline
\multirow{6}{*}{$0.1 < z < 0.2$} & 1 & $10.0 < \log\,[\mathcal{M}_{\ast}/\mathcal{M}_{\odot}] < 10.5$    & -- & -- & -- & -- & -- & -- \\[1pt] 
& 2 & $10.5 < \log\,[\mathcal{M}_{\ast}/\mathcal{M}_{\odot}] < 11.0$    & $0.16\pm0.08$ & $-2.98\pm0.07$ & 3.56 & 0.0961 & 0.31 & 10 \\[1pt] 
&  &  & [$0.25\pm0.05$] & [$-2.95\pm0.04$] & [23.39] & [{\color{blue} 0.0029}] & [0.80] & [8] \\[1pt] 
& 3 & $11.0 < \log\,[\mathcal{M}_{\ast}/\mathcal{M}_{\odot}] < 11.5$    & $0.35\pm0.05$ & $-3.27\pm0.05$ & 41.53 & {\color{blue} 0.0007} & 0.87 & 8 \\[1pt] 
&  &  & [$0.39\pm0.05$] & [$-3.24\pm0.04$] & [61.63] & [{\color{blue} 0.0005}] & [0.93] & [7] \\[1pt] 
& 4 & $\log\,[\mathcal{M}_{\ast}/\mathcal{M}_{\odot}] > 11.5$           & $0.77\pm0.08$ & $-3.32\pm0.07$ & 95.52 & {\color{blue} 0.0103} & 0.98 & 4 \\[1pt]  
  \hline 
 \end{tabular}
\end{table*}

Several works suggest a change in AGN X-ray luminosity (and BH accretion rate) depending on the position of the host galaxy on the SFR-$\mathcal{M}_{\ast}$ diagram with respect to the MS of SFG. The enhancement or suppression of X-ray luminosity (and BH accretion rate) for galaxies above the MS of SFG (i.e. starbursts) compared to the `normal' star-formation population of galaxies is rather controversial: according to some works, starbursts are less efficient in SMBH feeding than galaxies inside and below MS \citep{Masoura:18, Carraro:20}, while others support a simultaneous increase of the BHAR and SFR even for galaxies above the MS \citep{Pouliasis:22, Mountrichas:22} or the absence of correlation at all \citep{Rovilos:12, Shimizu:15}. 
 We highlight the position of the MS in Fig.\,\ref{fig:bhar-sfr-z1-z2} calculated using Eq.\,\eqref{eq:sfg-pass-line}. The position of the drop and the MS weakly correlate; however, it is hard to tell whether this is revealing an intrinsic physical dependence, especially considering that according to most studies, the MS flattens toward high stellar masses due to an increased fraction of bulge-dominated galaxies at higher $\mathcal{M}_{\ast}$ \citep{Erfanianfar:16, Tomczak:16, Schreiber:16, Popesso:19, Dimauro:22}. Thus the MS position in Fig.\,\ref{fig:bhar-sfr-z1-z2} may actually need to be shifted toward lower SFR relative to the observed sBHAR drop. Furthermore, the deficiency of high accreting AGN ($\log\,\lambda_{\rm sBHAR} > -2$) in our sample can also flatten our sBHAR-SFR relation because, according to low- and high-redshifts studies, quasars preferentially reside inside and above the MS (i.e. high SFR; \citealt{Pouliasis:22, Zhuang:22}). Finally, the flattening of the sBHAR-SFR relation toward high values of SFR may be also caused by deviations in the stellar-to-BH mass scaling relation, as discussed in the next Section.


\subsection{The relation between BH growth and SFR: comparison with the literature}\label{sec:disc-comparison}

The specific BH accretion rate defined in Eq.\,\eqref{eq:spec-bhar} is affected by uncertainties in the underlying BH-to-stellar mass scaling relation. A number of studies \citep{Reines:15, Savorgnan:16, Shankar:17, Shankar:20, Gonzalez:22, Graham:23a, Sahu:23} showed that the BH-to-stellar mass relation for local galaxies varies depending on morphology type. For instance, early-type  galaxies (i.e. spheroids or classical bulges) have a tendency to follow the canonical BH-to-bulge mass relation \citep{Haring:04, Kormendy:13, McConnel:13, LiJ:23}, which is considered to cause also the observed BH-to-stellar mass relation. At the same time,  late-type galaxies with pseudobulges usually show a weaker correlation between their BH mass and the host galaxy properties (like the mass of the pseudobulge or the disk component), and direct estimations of their BH masses suggest smaller values than those obtained from BH-to-stellar mass relation \citep{Shankar:16, LiJ:23}. This may result in an underestimation of the sBHAR (i.e. $\lambda_{\rm sBHAR} \propto L_{\rm X}/\mathcal{M}_{\ast}$) and may be responsible for the flattening of sBHAR-SFR relation at higher SFR. 

\subsubsection{Methodology}

To test the hypothesis {that the flattening of sBHAR-SFR relation at higher SFR is caused by uncertainties in the BH-to-stellar mass scaling relation} we study the absolute BH accretion rate ($\dot{m}_{\rm BH}$),  which represents the mass growth rate (in $\mathcal{M}_{\odot}$/year units) of the central BH, using the definition from \citealt{Alexander:12, Chen:13, Delvecchio:15, Rodighiero:15, Stemo:20}:
\begin{equation}
\dot{m}_{\rm BH}\,[\mathcal{M}_{\odot}\,{\rm year}^{-1}] = 0.15 \frac{\epsilon}{0.1} \cdot \frac{k_{\rm bol} L_{\rm X} [\text{erg\,s}^{-1}]}{10^{45}},   
\label{eq:m-dot}
\end{equation}
where $\epsilon$ is the mass-energy efficiency conversion (typically estimated to be $\epsilon \approx 0.1$, \citealt{Marconi:04}) and $k_{\rm bol} L_{\rm X} = L_{\rm bol}$ is the bolometric luminosity defined similarly to Section\,\ref{sec:bhar}. This also allows us to compare our results with the BHAR-SFR relations derived in the literature, for different AGN samples within the wide range of redshifts, without depending on the specifics of BH mass derivation. 

\begin{figure*}
\centering
\includegraphics[width=0.90\linewidth]{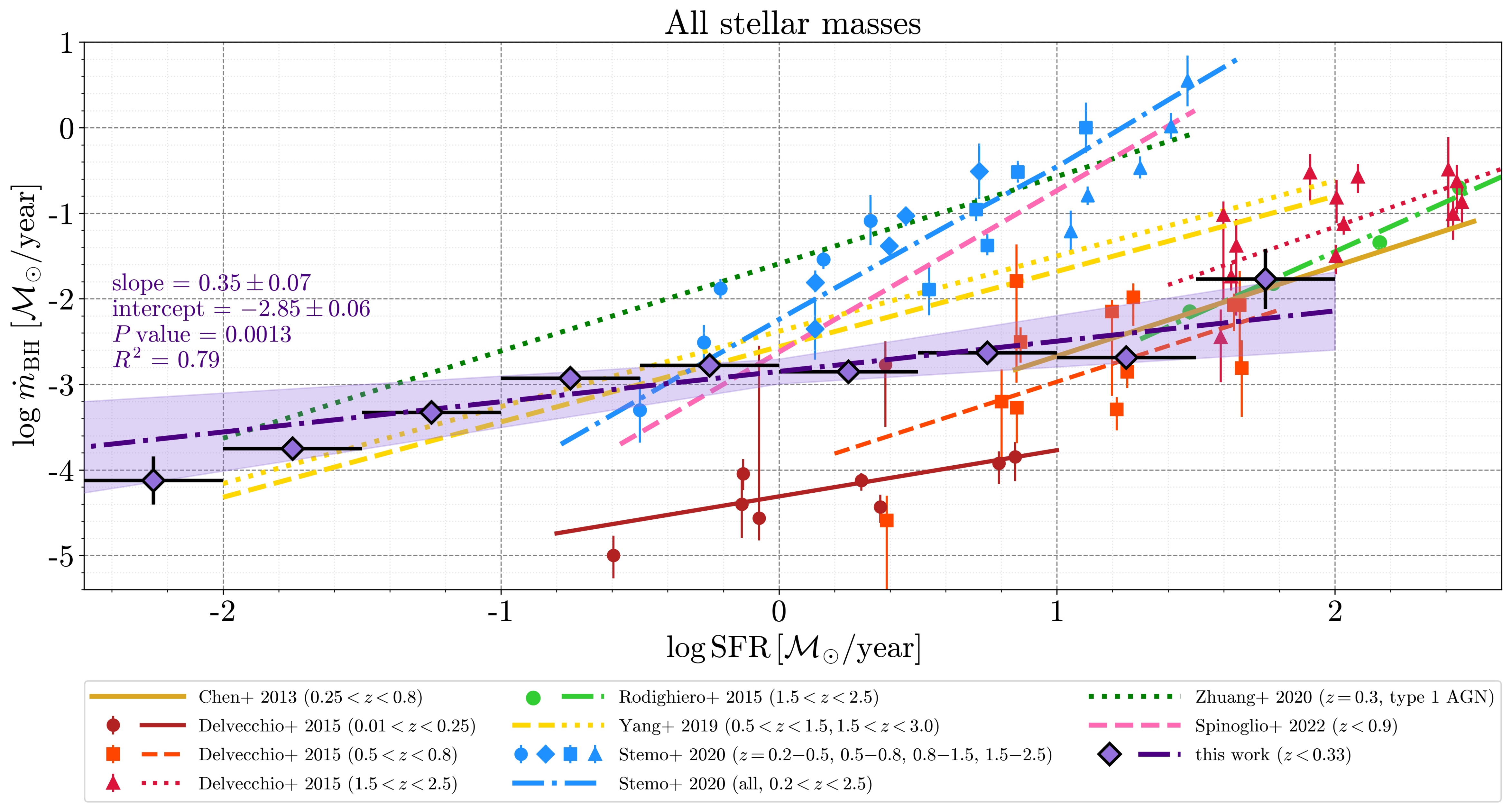}
\vspace*{-1.5ex}
\caption{The $\langle\log\,\dot{m}_{\rm BH}\rangle$--$\log\,{\rm SFR}$ relation for CSC+XMM sample estimated in the same way as sBHAR-SFR relation presented in Fig.\,\ref{fig:bhar-sfr-csc-xmm} together with results obtained by \citet{Chen:13, Delvecchio:15, Rodighiero:15, Yang:19, Stemo:20, Zhuang:20, Spinoglio:22} for AGN samples at different redshift intervals (see details in the text).}
\label{fig:bhar-sfr-others}
\end{figure*}

Following the same approach as presented in Section\,\ref{sec:bhar-sfr-chandra} we calculated the mean $\dot{m}_{\rm BH}$ in 10 bins of SFR in the range $-2.5 < \log\,{\rm SFR} < 2.0$ (with similar step in 0.5\,dex) for the entire CSC+XMM sample. In this case, we did not divide the sample into stellar mass or redshift ranges in order to facilitate comparison with results available in the literature. The uncertainties of each $\dot{m}_{\rm BH}$ were calculated similarly as in the previous section using jackknife resampling. The resulting relation and the best fitting parameters are presented in Fig.\,\ref{fig:bhar-sfr-others} together with the $\langle\log\,\dot{m}_{\rm BH}\rangle$--$\log\,{\rm SFR}$ relations from literature. 

\subsubsection{Consistency with results in the literature}

Figure\,\ref{fig:bhar-sfr-others} shows that $\langle\log\,\dot{m}_{\rm BH}\rangle$ correlates positively with $\log\,{\rm SFR}$ with a best-fit slope of $0.35\pm0.07$ ($P$-value $= 0.0013$), confirming the correlation between $\langle\log\,\lambda_{\rm sBHAR}\rangle$ and $\log\,{\rm SFR}$ found in the previous Section. Furthermore, even using $\langle\log\,\dot{m}_{\rm BH}\rangle$, we still observe that the relation between BH accretion rate and SFR flattens toward larger SFRs ($\log\,{\rm SFR} > 0$) compared to `quiescent' galaxies with $\log\,{\rm SFR} < 0$. 

The best-fit slope of our $\langle\log\,\dot{m}_{\rm BH}\rangle$--$\log\,{\rm SFR}$ relation is compatible with the one found in \citet{Delvecchio:15} at low-redshift ($0.01 < z < 0.25$), while the higher normalization is partly explained by the fact that their low redshift subsample contains galaxies with lower stellar masses ($\log\,[\mathcal{M}_{\ast}/\mathcal{M}_{\odot}] \lesssim 10.8$). At the same time, we observe that the slope of the low-redshift $\langle\log\,\dot{m}_{\rm BH}\rangle$--$\log\,{\rm SFR}$ relations   (both in \citealt{Delvecchio:15} and this work) are systematically flatter compared to high redshift studies \citep{Chen:13, Delvecchio:15, Rodighiero:15, Yang:19, Stemo:20}. This suggests that the correlation evolves with time, steepening at higher redshifts. However, we should also point out that those works do not sample well the low-SFR regime (i.e. they typically miss the quiescent galaxy population). In fact, the $\langle\log\,\dot{m}_{\rm BH}\rangle$--$\log\,{\rm SFR}$ relations found in \citet{Yang:19} for bulge-dominated systems (which are mainly located below the MS of SFG) for $0.5 < z < 1.5$ and $1.5 < z < 2.5$ reveal flatter slopes compared to the trends observed for the same intermediate-to-high redshift intervals in \citet{Chen:13, Delvecchio:15, Rodighiero:15}, and it agrees well with our local relation at $\log\,{\rm SFR} < 0$. In the case of \citet{Stemo:20}, they fit a large redshift range ($0.2 < z < 2.5$) and this makes it difficult to compare with  results derived on narrower redshift ranges; however at low redshift $0.2 < z < 0.5$ range (see blue circles in Fig.\,\ref{fig:bhar-sfr-others}) \citealt{Stemo:20} is in good agreement with our $\langle\log\,\dot{m}_{\rm BH}\rangle$--$\log\,{\rm SFR}$ relation. 

As mentioned before, our sample is missing highly accreting AGN (e.g. quasars) by construction, and the absence of such systems can be responsible (at least partially) for the flattening of the $\langle\log\,\dot{m}_{\rm BH}\rangle$--$\log\,{\rm SFR}$ relation. This is supported by the findings of \citet{Pouliasis:22, Zhuang:22} showing that quasars at low redshift are mainly located in the galaxies with high SFR. In fact, the $\dot{m}_{\rm BH}$-SFR relation derived by \citet{Zhuang:20} for the sample of type~1 AGN (i.e. $\log\,\dot{m}_{\rm BH} > -2$) at relatively low redshifts $z = 0.3$ shows a steeper slope compared to our local relation. A similar $\dot{m}_{\rm BH}$-SFR relation was presented by \citet{Spinoglio:22} for the combined sample of type~1 and type~2 AGN at $z < 0.9$, obtaining a best-fit slope steeper than other high redshift samples. 

\subsubsection{The role of stellar mass in triggering the BH growth and stellar formation}

\begin{figure}
\centering
\includegraphics[width=0.99\linewidth]{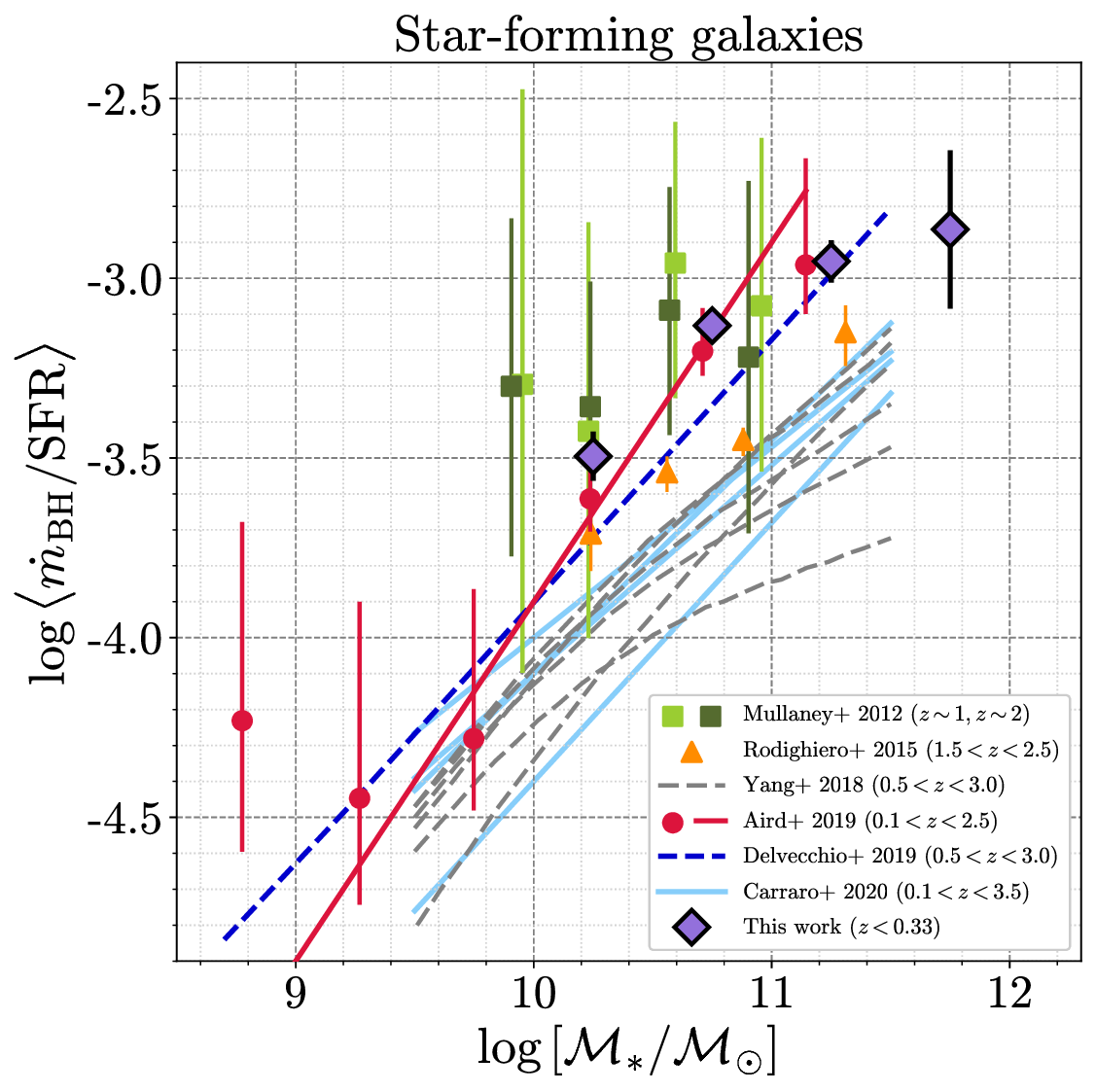}
\vspace*{-1.5ex}
\caption{The $\langle\dot{m}_{\rm BH}/{\rm SFR}\rangle$ as a function of stellar mass $\mathcal{M}_{\ast}$  for {\it star-forming galaxies} in our sample ($z < 0.33$) as well as results obtained in \citealt{Mullaney:12b, Rodighiero:15, Yang:18, Aird:19, Delvecchio:19, Carraro:20} (see description in the text).}
\label{fig:bharsfr-mass}
\end{figure}

As we saw above, although the average BH accretion rate correlates with SFR over a wide redshift range, the exact form of the correlation depends on the studied galaxy sample properties and the investigated redshift. In any case, this correlation does not clarify if there is a direct physical link (i.e. feedback) between these two processes or rather it arises from a common dependence on a more fundamental quantity, e.g. the amount of cold molecular gas in the host galaxy \citep{Aalto:12, Alexander:12, Davies:12, Combes:14, Sharon:16, Kakkad:17, Shimizu:19, Woo:20, Yesuf:20, Circosta:21, Koss:21, Zhuang:21, Salvestrini:22}. Taking into account the presence of the SFR--$\mathcal{M}_{\ast}$ correlation for the star-forming galaxies, it is uncertain whether the SFR or the stellar mass of the host galaxy plays a dominant role in triggering/regulating the SMBH growth \citep{Yang:18, Aird:19, Yang:19, Carraro:20}. To explore this point, in Fig.\,\ref{fig:bharsfr-mass} we derived the unitless quantity $\langle\dot{m}_{\rm BH}$/SFR$\rangle$ for star-forming galaxies, which represents the ratio of the mass accreted onto the SMBH relative to the mass accumulated into stars, in four stellar mass ranges. 

Figure\,\ref{fig:bharsfr-mass} shows that the ratios $\langle\dot{m}_{\rm BH}$/SFR$\rangle$ in SFG tend to increase with stellar mass indicating that, in comparison to low mass galaxies, more massive systems are more effective at feeding their central SMBH (and fuel faster SMBH growth) rather than forming stars. Such behaviour may be caused by more efficient transportation of cold gas toward the galaxy center, for instance, aided by the presence of a denser core in more massive galaxies (i.e. bulge, \citealt{Fang:13, Ni:21, Aird:22, Di:23}). The reduced accretion efficiency in lower mass galaxies may be also caused by the increased influence of the stellar feedback (e.g. supernova explosions), which reduces or interrupt the gas inflow to the central SMBH \citep{Fabian:12, Dubois:15, Hopkins:16, Emerick:18, Byrne:23}. 
Our measurements of $\langle\dot{m}_{\rm BH}$/SFR$\rangle$ are in good agreement with those present in the literature for different redshift ranges \citep{Mullaney:12b, Rodighiero:15, Aird:19, Delvecchio:19}. In this way, our result supports the scenario where the ratio between average black hole growth and average galaxy growth remains constant over cosmic times, despite the significant evolution of both the typical SMBH growth rates and star-formation rates over cosmic times \citep{Madau:14, Aird:15, Malefahlo:22}. An increase of $\langle\dot{m}_{\rm BH}$/SFR$\rangle$  with stellar mass was also found in \citet{Yang:18, Carraro:20} but their absolute values are systematically lower than ours, likely due to a different $k_{\rm bol}$ assumptions; in fact if we adopt a luminosity-dependent $k_{\rm bol}$ as they did, the two results are in close agreement. However, \citet{Yang:18} also suggest that $\langle\dot{m}_{\rm BH}/{\rm SFR}\rangle$ evolves with redshift up to $z < 3.0$. 

In contrast, we find that quiescent galaxies show no change of $\langle\dot{m}_{\rm BH}$/SFR$\rangle$ with stellar mass, which agrees with findings in \citealt{Carraro:20}. Actually, quiescent galaxies have a higher level of $\log\,\langle\dot{m}_{\rm BH}$/SFR$\rangle$ (near $-2.08$ on average\footnote{Note that we do not show $\langle\dot{m}_{\rm BH}$/SFR$\rangle$ points for quiescent galaxies in Fig.\,\ref{fig:bharsfr-mass} to avoid confusion with the results for SFG.}) indicating that they are comparatively more efficient in feeding the SMBH than in forming stars. Moreover, this is pointing toward the existence of the different physical mechanisms responsible for AGN fuelling in quiescent galaxies via stellar mass-loss or cold accretion flows \citep{Rafferty:06, Kauffmann:09, Woodrum:22, Bambic:23, Guo:23} and/or different accretion mode (e.g. the convection/advection-dominated accretion flows, Bondi accretion of hot gas; \citealt{Narayan:97, Quataert:00, Allen:06, Hardcastle:07, Russell:13}). 

\section{Summary and conclusions}\label{sec:concl}

This paper presents a study of the correlation between AGN activity and stellar formation in the nearby Universe, improving and extending the analysis performed in \citetalias{Torbaniuk:21}. We started from the same parent galaxy sample extracted from the SDSS, contained in the {\it galSpec} catalogue, which provides spectroscopical estimates of SFR and $\mathcal{M}_{\ast}$ for each galaxy. In order to quantify the nuclear activity, we combined X-ray data from the \textit{Chandra} Source Catalog 2.0, with the 3XMM-DR8 data previously used in \citetalias{Torbaniuk:21}. 
This allowed us to: i) increase the AGN sample size, deriving more robust constraints on the specific BH accretion rate  distribution in the local Universe, $p(\log\,\lambda_{\rm sBHAR}|\mathcal{M}_{\ast}),$ as well as on the correlation between SFR and AGN activity, ii) adopt more stringent selection criteria to avoid mass-related biases and iii) demonstrate that the resolution limit of {\it XMM-Newton} was not significantly affecting our previous results.

We found that $p(\log\,\lambda_{\rm sBHAR}|\mathcal{M}_{\ast})$ has an approximately power-law shape, flattening or declining toward lower sBHAR ($\log\,\lambda_{\rm sBHAR}\lesssim -3.5$) for a wide range of stellar masses, supporting a picture where the local Universe contains predominately SMBHs accreting at low efficiency with respect to earlier epochs. Furthermore, the fraction of `classical' AGN with high-efficient accretion ($\log\,\lambda_{\rm sBHAR} > -2.0$) reaches only 0.4~per\,cent relative to 7--24 per\,cent of the `entire' local AGN population with $\log\,\lambda_{\rm sBHAR} > -5.0$. At the same time, star-forming galaxies show generally higher AGN fraction (up to 47~per\,cent) compared to quiescent galaxies (up to 22~per\,cent).

We investigated the correlation between AGN activity and host galaxy properties, such as stellar mass and SFR, confirming that the nuclear X-ray luminosity depends on the stellar mass of the host galaxy for both star-forming and quiescent systems, while the median $\log\,\lambda_{\mathrm{sBHAR}}$ (i.e. $L_X$ normalised to the host galaxy/SMBH mass) seems independent from the stellar mass for SFG and is possibly anticorrelated for quiescent galaxies. Additionally, quiescent galaxies show systematically lower X-ray luminosity and $\lambda_{\rm sBHAR}$ at a given stellar mass with respect to the star-forming galaxies, implying a significant correlation between $\log\,\lambda_{\mathrm{sBHAR}}$ and $\log$\,SFR for almost all stellar masses. We discuss the difficulties in comparing studies at different redshifts, due to the different $L_X$, mass, and SFR ranges that they probe. In general, however, we observe a weaker dependence of the absolute SMBH accretion rate on SFR, in the sense that our best-fit relation appears flatter, suggesting a smaller amount of accreting material in low-redshift galaxies. 

At the same time, we found that the SMBH accretion rate relative to the mass formed into stars (for star-forming galaxies), i.e. $\langle\dot{m}_{\rm BH}/{\rm SFR}\rangle$, increases with stellar mass, which is fully consistent with results at high redshifts. This may indicate that the rate at which both AGN and star-formation are triggered in star-forming galaxies primarily depends on the total stellar mass. On the other end, the absence of $\langle\dot{m}_{\rm BH}/{\rm SFR}\rangle$--$\mathcal{M}_{\ast}$ correlation for quiescent galaxies suggests that a different physical mechanism is responsible for the triggering and fuelling AGNs in quiescent galaxies. 

Our results can be further validated through the studies of the cold molecular gas content in the host galaxies and its correlation with the properties of AGN and SF processes. However, currently available studies are limited to the small samples of mainly local galaxies and present rather contradictory results showing an enhanced/depleted (or no difference) fraction of molecular gas in AGN relative to non-AGN host galaxies. Thus, also points to the need for a more systematic approach to searching for cold molecular gas for larger samples of galaxies and AGN.


\section*{Acknowledgements}

Authors are grateful to Ivan Delvecchio for helpful discussions and for providing his relevant data for proper comparison with our results. Also, we thank Jarle Brinchmann for his help with understanding the data presented in {\it galSpec} catalogue. 

OT and MP acknowledge the financial contribution from the agreement ASI-INAF n.2017-14-H.O. and from PRIN-MIUR 2022. MP also acknowledges support from the Timedomes grant within the ``INAF 2023 Finanziamento della Ricerca Fondamentale''. AG~acknowledges support from the EU H2020-MSCA-ITN-2019 Project 860744 ``BiD4BESt: Big Data applications for Black hole Evolution Studies''\footnote{\url{www.bid4best.org}} and the Hellenic Foundation for Research and Innovation (HFRI) project ``4MOVE-U'' grant agreement 2688, which is part of the programme ``2nd Call for HFRI Research Projects to support Faculty Members and Researchers''.
FJC acknowledges support from the grant PID2021-122955OB-C41 funded by MCIN/AEI/10.13039/501100011033 and by ERDF A way of making Europe.


Funding for SDSS-III has been provided by the Alfred P. Sloan Foundation, the Participating Institutions, the National Science Foundation, and the U.S. Department of Energy Office of Science. The SDSS-III web site is \url{http://www.sdss3.org/}. 

This research has made use of data obtained from the {\it Chandra} Source Catalog, provided by the {\it Chandra} X-ray Center (CXC) as part of the {\it Chandra} Data Archive and from the 3XMM-Newton serendipitous source catalogue compiled by the 10 institutes of the {\it XMM-Newton} Survey Science Centre selected by ESA.

This publication makes use of data products from the Two Micron All Sky Survey, which is a joint project of the University of Massachusetts and the Infrared Processing and Analysis Center/California Institute of Technology, funded by the National Aeronautics and Space Administration and the National Science Foundation.

\section*{Data availability}

All the data presented/used in this work are publicly available. The SDSS data (\textit{galSpec} catalogue) are accessible through the online service \texttt{CasJobs} SDSS SkyServer or the web-page \url{https://www.sdss.org/dr12/spectro/galaxy\_mpajhu/}. The \textit{XMM-Newton} data are available in the \textit{XMM-Newton} Survey Science Centre (\url{http://xmmssc.irap.omp.eu/Catalogue/3XMM-DR8/3XMM_DR8.html}).  The count/flux upper limits for XMM data are obtained from the XMM FLIX at \url{https://www.ledas.ac.uk/flix/flix.html}. The {\it Chandra} data is available to access through the official {\tt CSCview} application (see \url{https://cxc.harvard.edu/csc/gui/intro.html}). The 2MASS data can be found through the Infrared Processing \& Analysis Center (\url{https://old.ipac.caltech.edu/2mass/}).

\bibliographystyle{mnras}
\bibliography{references} 



\bsp	
\label{lastpage}
\end{document}